\documentclass[pra,twocolumn,pra,groupedaddress]{revtex4-1}

\usepackage{graphicx}
\usepackage{dcolumn}
\usepackage{bm}
\usepackage{epsfig}
\usepackage{amsmath}
\usepackage{amssymb}
\usepackage{color}
\usepackage{bbm}
\usepackage{braket}
\usepackage{float}

\usepackage[colorlinks=true,citecolor=blue,linkcolor=blue,urlcolor=blue]{hyperref}

\begin{document}

\title{Magnetic-field-induced cavity protection for intersubband polaritons}

\author{Daniele De Bernardis$^1$, Mathieu Jeannin$^2$, Jean-Michel Manceau$^2$, Raffaele Colombelli$^2$, Alessandro Tredicucci$^3$, and Iacopo Carusotto$^1$}
\affiliation{$^1$INO-CNR BEC Center and Dipartimento di Fisica, Universit{\`a} di Trento, 38123 Povo, Italy}
\affiliation{$^2$Centre de Nanosciences et de Nanotechnologies (C2N), CNRS UMR, 9001, Université Paris-Saclay, Palaiseau 91120, France}
\affiliation{$^3$NEST, CNR-Istituto Nanoscienze, and Dipartimento di Fisica “E. Fermi”, Università di Pisa, Largo Pontecorvo 3, 56127 Pisa}

\date{\today}

\begin{abstract} 
We analyse the effect of a strong perpendicular magnetic field on an intersubband transition in a disordered doped quantum well strongly coupled to an optical cavity. 
The magnetic field changes the lineshape of the intersubband optical transition due to the interface roughness of the quantum well from a Lorentzian to a Gaussian one. In this regime, a novel form of magnetic-field-induced cavity protection sets in, which strongly reduces the polariton linewidth to the cavity contribution only. Implications of our results for fundamental studies of nonlinear polariton dynamics and for technological applications to polariton lasers are finally highlighted.
\end{abstract}
 
\maketitle

%
%


\maketitle

Strongly confined quasi two-dimensional (2D) electronic systems have a long research history \cite{Ando_RevModPhys.54.437}, and still remain under the researchers' spotlight due to their large importance for both fundamental science and technological developments.
In recent years, 2D electronic systems supporting \emph{intersubband (ISB) transitions} polarized perpendicular to the quantum well plane have been combined with good-quality electromagnetic cavities giving rise to strong light-matter coupling effects and opening the field of intersubband polaritons \cite{Tredicucci_PhysRevLett.90.116401, Polariton_panorama}.

In particular, by adjusting the electronic density it is possible to vary the dipole moment of the ISB on a quite broad range, bringing these polaritonic devices from the usual strong-coupling regime to the more exotic \emph{ultra-strong coupling} (USC) regime. This regime was experimentally achieved with ISB and other various solid-state platforms~\cite{Dietze_USC_doi:10.1063/1.4830092,Sirtori_USC_MIR_doi:10.1063/1.3598432,Askenazi_2014,Todorov_PhysRevLett.105.196402}  and is of great interest for its strongly modified ground-state properties~\cite{Ciuti_PhysRevB.72.115303,nori_USC_review, Solano_USC_RevModPhys.91.025005}. More in general, the large flexibility of ISB polaritons holds great promise for the study of a wide range of cavity quantum electrodynamics (cQED) effects from both a fundamental \cite{knorr2022intersubband, Ciuti_PhysRevB.72.115303, Ciuti_DeLiberato_PhysRevB.87.235322, Ciuti_NatPhot_nonAdiabSwitchCavityQED} and an applied perspective~\cite{Colombelli_2005, Vasanelli_PhysRevLett.100.136806, DeLiberato_Ciuti_PhysRevLett.102.136403}.
Despite the great achievements reached using ISB polaritonic devices in both the mid-infrared and THz regime, a serious obstacle hindering a full development of this research program is caused by the relatively large linewidth of the ISB transition~\cite{Colombelli_Perspectives_PhysRevX.5.011031}, mostly due to interface roughness of the semiconductor quantum well nanostructures~\cite{Unuma_doi:10.1063/1.1535733}. 

In this article we theoretically investigate a method to dramatically reduce the ISB polariton linewidth by suppressing the contribution of interface roughness and  bringing it down to the cavity contribution only. The idea is to combine a strong light-matter coupling to a single electromagnetic cavity mode with a strong static magnetic field, so as to enter a regime of {\em cavity protection}~\cite{FirstCavityProtection_Houdre_PhysRevA.53.2711}. The strong perpendicular magnetic field quenches the in-plane kinetic energy of the electrons and strongly localises them in the disorder potential due to the interface roughness of the quantum well. As a consequence, the line shape of the ISB transition changes from a standard Lorentzian shape to an almost Gaussian one. Due to the fast decay of the Gaussian tails, the strong coupling to the cavity mode then effectively suppresses the inhomogenous broadening of the ISB transition, leaving a polariton linewidth which is then only limited by the cavity component. 
Differently from previous works on cavity protection in multi-quantum-well devices~\cite{Colombelli_immunity_PhysRevB.96.235301} and in various ensembles of emitters in cavity-QED devices~\cite{Molmer_CavityProtection_PhysRevA.83.053852, Diniz_CavityProtection_PhysRevA.84.063810, Mayer_CavityProtection_cQED_exp_NaturePhys, Faraon_cavity_protection_exp_nanoph_cavity, Breeze_cavity_protection_molecules, Long_arxiv_cavityProtection2022_https://doi.org/10.48550/arxiv.2208.12088}, the magnetic-field-induced cavity protection mechanism predicted in our work is based on the in-plane dynamics of electrons in a single quantum well, which completely changes nature under the effect of the strong magnetic field. Our calculations anticipate a dramatic enhancement of the polariton quality factor, which opens new perspectives to experimental research in this field and to technological applications.


The article is organised as follows.
In section \ref{sec:model} we introduce the model used to describe ISB polaritons in the presence of interface roughness in the quantum well and calculate the cavity transmittivity in the strong coupling regime.
In section \ref{sec:Linewidth broadening due to roughness disorder} we characterise the impact of disorder on the ISB linewidth in the absence of a surrounding cavity and we extend the known results to the case of a strong perpendicular magnetic field.
In section \ref{sec:Intersubband polaritons and cavity protection} we translate the concept of cavity protection to our context of ISB polaritons in strong magnetic field and we show its implications: the main regimes are characterized and experimental implementations are discussed.
In section \ref{sec:conclusions} we summarize the main results and conclusions of this work. Additional information on the derivations are given in the Appendices.

\section{Model}
\label{sec:model}

\begin{figure}
\centering
	\includegraphics[width=\columnwidth]{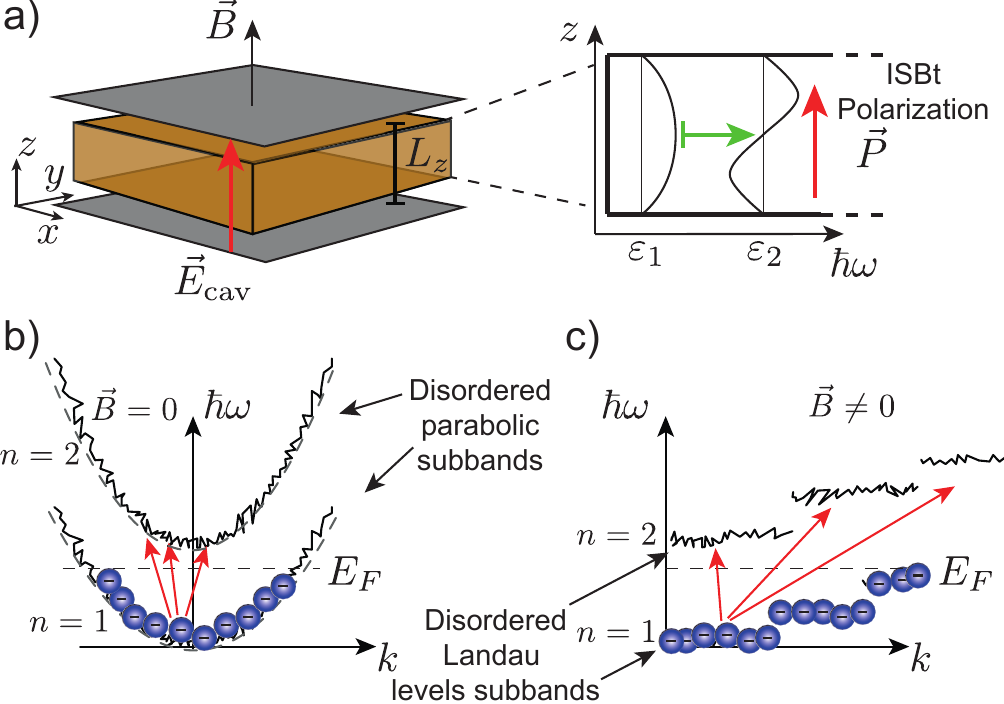}
	\caption{ a) Sketch of the system. A semiconductor two-dimensional quantum well is inserted in a metallic cavity. The electric field of the cavity mode oscillates along the $z$-direction, perpendicular to the quantum well plane. The electronic intersubband transitions in the quantum well give rise to an electronic polarization directed along the same direction of the cavity electric field, realising a strong light-matter coupling regime between the cavity and the quantum well. b-c) Schematic view of the electronic subbands in the disordered quantum well in absence/presence of the external magnetic field perpendicular to the quantum well plane. The lowest optically-active excitations are transitions between the lowest two subbands due to the electrons below the Fermi energy $E_F$. Differently from the clean case, these transitions are not only fully vertical and each electron can jump to different eigenstates of the upper subband. }
	\label{fig:fig1}
\end{figure}

In this section we introduce the model to describe a planar quantum well (QW) coupled to the electromagnetic field of an optical cavity, as represented in Fig. \ref{fig:fig1}(a). Electron motion along the quantum well plane is typically affected by interface roughness disorder, caused by the fabrication process. The inclusion of this disorder forces us to go beyond the standard theory of intersubband transitions based on the bosonization of the collective ISB electronic excitations~\cite{Todorov_Sirtori_PhysRevB.85.045304} and include the underlying fermionic degrees of freedom. However in the small disorder, small excitation limit, it is still possible to derive a sufficiently simple and manageable theory of light-matter interactions with ISB transitions. This theory captures well the effect of disorder and can be immediately extended with the inclusion of a static external magnetic field orthogonal to the QW. 

\subsection{Electronic states in a disordered quantum well in the presence of a static magnetic field }

We consider here an electronic system strongly confined in the $z$ direction and free in the $(x,y)$-plane, in such a way to realise a two dimensional quantum well (QW) with well defined energy subbands \cite{Capasso_liu2000intersubband}.
An homogeneous magnetic field with amplitude $B$ is pointing in the $z$-direction, perpendicular to the two dimensional QW.
Assuming the symmetric gauge for the magnetic vector potential, $\vec{A}(\vec{r}) = B/2 (-y,x, 0)$, the single particle Hamiltonian for each electron reads
\begin{equation}
H_e = \frac{\left(\vec{p} - e\vec{A} \right)^{\, 2}}{2m^*} + \frac{p_z^2}{2m^*} + U( z, \vec{r}),
\end{equation}
where $\vec{p}=(p_x, p_y)$, $\vec{r}=(x,y)$ are respectively the electron's in-plane momentum and position, $m^*$ is the electron effective mass, $e$ its charge and $U( z, \vec{r})$ is the single particle confining potential.
Under the assumption of a strong confinement on the $z$-direction we can employ a Born-Oppenheimer-like approximation between the different directions, and write the single particle eigenfunctions as
\begin{equation}\label{eq:eigenfun_general_BO}
\psi(z, \vec{r}) \approx \zeta_n(z, \vec{r}) \varphi_{n \,k}(\vec{r}),
\end{equation}
where the subband wavefunctions $\zeta_n(z, \vec{r})$ are assumed to depend only parametrically on the in-plane position $\vec{r}$ and  for each $\vec{r}$ solve the eigenproblem
\begin{equation}
\left[ \frac{p_z^2}{2m^*} + U( z, \vec{r}) \right]\zeta_n(z, \vec{r}) = \varepsilon_n(\vec{r}) \zeta_n(z, \vec{r})
\end{equation}
along the $z$ direction. Here, $n=1,2 \ldots$ are integer numbers and $\varepsilon_n(\vec{r})$ indicate the Born-Oppenheimer potential energy interfaces.

For each subband $n$, the in-plane eigenfunctions $\varphi_{n \, k}(\vec{r})$ are thus obtained by diagonalising the in-plane Hamiltonian 
\begin{equation}
H_n = \frac{\left(\vec{p} - e\vec{A} \right)^{\, 2}}{2m^*} + \varepsilon_n(\vec{r})
\end{equation}
and are labeled generically by the quantum number $k$ (or array of numbers). In the case of a clean QW with no magnetic field $\vec{A}=0$, and no spatial dependence of the potential interfaces $\varepsilon_n(\vec{r} ) = \varepsilon_n$, this index is actually a pair of numbers representing the wavevector of the planewave basis. In spatially inhomogenoeus configurations, where translational invariance is broken and $\varepsilon_n(\vec{r} )$ depends on the position, planewaves are no longer eigenstates of the system. In this case the index $k$ represent a generic quantum number labeling the true eigenstates of the system, extracted by exactly diagonalising the full Hamiltonian. 

We assume now that the $z$-confinement is characterised by a single length scale  $L_z = L_z(\vec{r})$ which depends on the in-plane position $\vec{r}$, and that the potential energy interface depends from the in-plane position only through this characteristic length $\varepsilon_n(\vec{r}) = \varepsilon_n(L_z(\vec{r}))$.
Moreover we assume that $L_z(\vec{r})$ is given by a fixed length over the whole plane plus a small fluctuating part $L_z(\vec{r}) = L_{\rm qw} + \delta L(\vec{r})$ which describes the roughness of the QW interface.
Expanding the potential energy interface at the first order in $\delta L$ we arrive to the final expression of the in-plane electronic Hamiltonian of each subband as $H_n \approx H_n^{\parallel} + \varepsilon_n^{\rm qw}$, where the in-plane electronic Hamiltonian is given by
\begin{equation}\label{eq:Ham_in-plane_generic}
H_n^{\parallel} = \frac{\left(\vec{p} - e\vec{A} \right)^{\, 2}}{2m^*} + \delta U_n(\vec{r})~,
\end{equation}
$\varepsilon_n^{\rm qw} = \varepsilon_n (L_{\rm qw})$ is the energy of the $n$ subband of the clean QW, and $\delta U_n(\vec{r}) = \partial_L \varepsilon_n(L_z) \cdot \delta L(\vec{r})$ is the position-dependent energy shift due to the disorder.

The effect of the external magnetic field $B$ on the energy distribution of the in-plane eigenstates is sketched in Fig.~\ref{fig:fig1}(b,c): when the cyclotron frequency $\omega_B = eB/m^*$ is larger than the typical energy width of each potential energy interface $\delta U_n$, the electronic dispersion of each subband changes from the usual parabola to discrete Landau levels.

In our description we completely neglect the effect of electron-electron interactions. These are known to have a small effect in the ISB optical linewidth \cite{Luin_PhysRevB.64.041306}, which is mainly limited by the interface roughness \cite{Nikonov_PhysRevLett.79.4633,Vignale_PhysRevLett.87.037402, Unuma_doi:10.1063/1.1535733}. For this reason, we chose to avoid further complications and focus on the single particle dynamics only, so to highlight in the most transparent way the interplay between roughness disorder and magnetic field. Including the effect of Coulomb interactions in our description of the in-plane electron dynamics goes beyond the scope of this paper and is left for a future work. 

In spite of this, it must be kept in mind that the effect of electron-electron interactions is still present in the presence of magnetic field \cite{Unterrainer_PhysRevLett.88.226803,Kempa_PhysRevB.68.085302} and, in particular, the effect of Coulomb interactions onto the electron motion along the $z$-axis is completely unaffected by the magnetic field along $z$. As such, it keeps giving an important contribution to the nonlinearity of the ISB response to strong electromagnetic fields~\cite{cominotti2021theory}. This implies that our results are directly applicable to the on-going quest of developing ISB polaritonic devices with strong optical non-linearities and polariton lasers, where a reduced polariton linewidth can be a game-changing improvement~\cite{knorr2022intersubband}.

\subsection{Intersubband transitions and cavity-polaritons}

After having characterized the single-particle eigenstates of electrons, we consider here that the two dimensional QW is placed inside a cavity enclosed within a pair of metallic plates parallel to the QW $(x,y)$-plane, as represented in Fig. \ref{fig:fig1}. We assume that the electronic ISB transitions mostly couple to a single TM-polarized cavity mode, which, in the long-wavelength approximation, can be taken to be homogeneous along the plane, with frequency $\omega_c$ and zero-field amplitude $E_0=\sqrt{\hbar \omega_c/(2\epsilon_0 V)}$ directed in the $z$-direction and determined by the cavity volume $V$. 

The light-matter Hamiltonian is then given by the dipole-gauge Hamiltonian \cite{Todorov_Sirtori_PhysRevB.85.045304}, 
\begin{equation}
H_{\rm int} \approx \int dz d^2 r \vec{P}(\vec{r}, z) \cdot \vec{E}_{\rm cav}(\vec{r}, z)\,,
\end{equation}
where $\vec{P}$ is the electronic polarization of the QW and $\vec{E}_{\rm cav} \approx \vec{u}_z E_0 (a + a^{\dag})$ is the electric field of the cavity.
Assuming we are below the ultra-strong light-matter coupling regime, we can for simplicity discard the $\sim \vec{P}^2$-term~\cite{DeBernardis_PhysRevA.97.043820, Rubio_Rokaj_2018}.
Under this assumptions we can also neglect higher-order boundary effects, such as the presence of image charges on the plates, which are only manifested in the ultra-strong coupling regime and they only provided a renormalization of the overall frequency scales \cite{Todorov_PhysRevB.91.125409, DeBernardis_PhysRevA.97.043820}.
In this respect we can argue that the thickness of the cavity and the distance between the QW and the metallic plates does not really play any role.
This is specially the case of cavities used for ISB polaritons where the field is mostly homogeneous along the growth direction \cite{Todorov:10}. For these reasons the size of the cavity will not enter as a physical parameter in our description.

Introducing the creation/annihilation spinless electronic operators $\Psi(\vec{r}, z)$, satisfying the Fermionic anti-commutation relation $ \lbrace{ \Psi(\vec{r}, z)^{\dag}, \Psi(\vec{r}{\, ' }, z') \rbrace} = \delta(z-z') \delta^{(2)}(\vec{r}-\vec{r}\, ')$ we can rewrite the electronic polarization as $\vec{P}(\vec{r}, z) \approx \, e z\Psi^{\dag}(\vec{r}, z)\Psi(\vec{r}, z) \vec{u}_z $ \cite{Todorov_Sirtori_PhysRevB.85.045304}, where $e$ is the electron charge. The second quantized light-matter coupled Hamiltonian reads
\begin{equation}
\begin{split}
H_{\rm tot} & = \hbar \omega_c a^{\dag} a + \int dz\, d^2r  \, \Psi^{\dag}(\vec{r}, z) H_e \Psi (\vec{r}, z) +
\\
& + e \int dz\, d^2r  \, z \Psi^{\dag}(\vec{r}, z)\Psi(\vec{r}, z) \cdot E_0 (a + a^{\dag}).
\end{split}
\end{equation}
Restricting ourselves to QWs with a fixed number of electrons $N_e$, we define the \emph{plasma frequency} of the electronic transitions as
\begin{equation}
\omega_{\rm P} = \sqrt{\frac{e^2}{\epsilon_0 m^*} \frac{N_e}{V}}.
\end{equation}
The lowest QW intersubband frequency transition and its $z$-direction oscillator strength are defined as
\begin{align}\label{eq:omega_QW_f_QW}
\hbar \omega_{\rm qw} = \varepsilon_2^{\rm qw} - \varepsilon_1^{\rm qw} && f_{\rm qw} = \frac{2 m^* \omega_{\rm qw} z_{21}^2}{\hbar}\,,
\end{align}
in terms of the dipole matrix element $z_{n\, n'}^2 = |\braket{\zeta_n | z | \zeta_{n'}} |^2$ between the $n,n'$ single particle eigenfunctions defined in Eq. \eqref{eq:eigenfun_general_BO}. To avoid complications stemming from the implicit parametric dependence of the matrix elements $z_{n\, n'} \sim L_{\rm qw} + \delta L(\vec{r})$ on the in-plane position $\vec{r}$, we consider an average value $z_{n\, n'}^2$ over the $(x,y)$-plane, neglecting its dependence on the fluctuating QW thickness. This assumption is well justified within the weak disorder assumption and corrections are of higher order in $\delta L(\vec{r})$.

We then re-express the electron field operator in terms of the single particle eigenfunctions $\Psi (\vec{r}, z) = \sum_{n \, k} \zeta_n(z, \vec{r}) \varphi_{n \, k}(\vec{r}) c_{n \, k}$, where the creation operators $c_{n \, k}$  labelled by the subband $n$ and in-plane $k$ indices satisfy Fermionic commutation rules $\lbrace{ c_{n \, k}, c^{\dag}_{n' \, k'} \rbrace} = \delta_{n, n'} \delta_{k, k'}$. 
Within a rotating wave approximation, we neglect the counter-rotating terms in the light-matter interaction and we restrict our attention to the lowest intersubband transition between $n=1$ and $n=2$. In this way, we obtain the \emph{cavity-plasma} Hamiltonian
\begin{equation}\label{eq:Ham_cavity_plasma}
\begin{split}
H_{\rm tot} & \approx \hbar \omega_c a^{\dag} a + \sum_{n=1,2 \, k} \left( \varepsilon_n^{\rm qw} + \hbar\omega_n^{\parallel}(k) \right) c^{\dag}_{n \, k} c_{n \, k}
\\
& + \frac{\hbar \Omega_R}{2}\left( a \cdot \sum_{k,k'} \Lambda_{k \, k'}   c^{\dag}_{2\, k} c_{1 \, k'} + {\rm h.c.} \right)\, ,
\end{split}
\end{equation}
where the strength of the light-matter coupling is quantified by the \emph{Rabi frequency}
\begin{equation}
    \Omega_R = \omega_{\rm P} \sqrt{f_{\rm qw} \frac{\omega_c}{\omega_{\rm qw}} }\,.
\end{equation}
Here, $\omega_n^{\parallel}(k)$ are the eigenfrequencies of $H_n^{\parallel}$ as defined in Eq. \eqref{eq:Ham_in-plane_generic} and the $k,k'$ transition matrix element is given by the in-plane wavefunction overlap between states in the two different subbands
\begin{equation}
\Lambda_{k \, k'} = \frac{\braket{\varphi_{2\, k} | \varphi_{1\, k'}}}{\sqrt{N_e}}.
\end{equation}
Note that the use of the dipole gauge allows us to truncate the Hilbert space to the two lowest subbands without introducing potentially dangerous spurious effects \cite{DeBernardis_PhysRevA.98.053819}.

\subsection{Cavity transmission and quantum well optical density}

Assuming that the cavity is resonant with the lowest QW intersubband transition $\omega_c \sim \omega_{\rm qw}$, we can use the cavity-plasma Hamiltonian in Eq. \eqref{eq:Ham_cavity_plasma} to derive the input-output equations for the system \cite{gardiner00} and then the cavity optical transmission of a weak probe of frequency $\omega$ at normal incidence (see App. \ref{app:sec:in_out_th} for further details on the derivation ), 
\begin{equation}\label{eq:cavity_transmission}
T_c(\omega ) = - \frac{\gamma_c/2}{\omega - \omega_c + i \frac{\gamma_c}{2} + \frac{\Omega_R^2}{4}\chi_{\rm qw}(\omega) },
\end{equation}
where $\gamma_c$ is the cavity loss rate and microscopic details of the QW are summarized by the optical response
\begin{equation}\label{eq:electronic_Susceptibility_optical_response}
\chi_{\rm qw}(\omega ) =  - \sum_{k'\leq k_F', k} \frac{\left| \Lambda_{k, k'} \right|^2}{\omega - \omega_{\rm qw} - \omega_{k \, k'}^{\parallel} + i\kappa }\,.
\end{equation}
Here $\omega_{k \, k'}^{\parallel} + \omega_{\rm qw} = \omega_2^{\parallel}(k) - \omega_1^{\parallel}(k') + \omega_{\rm qw}$ is the frequency of the $(1,k')\rightarrow (2,k)$ transition, $\kappa$ is a small phenomenological energy loss rate introduced to regularize the response, and the sum over initial states $k'$ is restricted to the occupied states below the Fermi level $k_F'$ of the lowest subband.

Using the Sokhatsky identity $\lim_{\kappa \rightarrow 0} {\rm Im}\left[ 1/(\omega + i\kappa) \right] = -\pi \delta(\omega)$ we can introduce the so-called \emph{QW optical density}
\begin{equation}\label{eq:QW_optical_density}
\begin{split}
\rho_{\rm qw} (\omega ) &= \lim_{\kappa \rightarrow 0} \frac{1}{\pi} {\rm Im}\left[ \chi_{\rm qw}(\omega ) \right]
\\
& = \sum_{k'\leq k_F', k} \left| \Lambda_{k, k'} \right|^2 \delta (\omega - \omega_{\rm qw} - \omega_{k \, k'}^{\parallel}).
\end{split}
\end{equation}
This quantity is normalised to one  $\int d\omega\, \rho_{\rm qw} (\omega ) = 1$, as can be verified by considering that $\mathbf{1} = \sum_{k} | \varphi_{n \, k} \rangle \langle \varphi_{n \, k} |$ and using again the Sokhatsky identity, and is equivalent to the density of states typically used in the literature~\cite{Ando_ISB_space_charge_layers}, which is derived from the electronic Green's function, as it is shown in App. \ref{app:sec:Green_fun}.
The QW optical response can be rewritten only in terms of the optical spectral density as
\begin{equation}
\chi_{\rm qw}(\omega ) = - \int d \omega ' \frac{\rho_{\rm qw} (\omega ' )}{\omega - \omega ' + i \kappa},
\end{equation}
making this quantity the central object of our investigation.

\section{Linewidth broadening due to roughness disorder}
\label{sec:Linewidth broadening due to roughness disorder}
To understand the behaviour of the cavity trasmission $T_c(\omega )$ in the different regime, we first need a clear understanding of the QW optical response $\chi_{\rm qw}(\omega )$, which represents the optical susceptibility of the QW to an external probe with incident frequency $\omega$. This reduces to study how the optical density $\rho_{\rm qw}(\omega )$ is modified by the disorder and the external magnetic field.

In a clean sample, or when the disorder is exactly the same in the two subbands the transition matrix is $\Lambda_{k \, k'} = \delta_{k\, k'}$ and all transition frequencies coincide, so to have $\omega_{k \, k'}^{\parallel} = 0$.
From Eq. \eqref{eq:QW_optical_density}, it then follows that the QW optical density is a delta function centered at the QW frequency $\rho_{\rm qw} (\omega ) \sim \delta (\omega - \omega_{\rm qw})$. 

In this section we will examine how this picture is destroyed by the disorder and, successively, what is the interplay between disorder and magnetic field.  To be concrete we keep the discussion in the simple, paradigmatic, example of the QW with infinite well confinement.
However, in order to make our results general and independent from the precise shape of the confining potential in the $z$-direction, we will introduce a set of natural units to express all the quantities in adimensional form. In this way all the results of this section applies as well to any other type of $z$-confinement after providing the appropriate rescaling of variables.

\subsection{Simple model for the interface roughness disorder}

When the system is instead affected by the interface roughness giving a different disorder potential $\delta U_n$ for each subband, several $k'\rightarrow k$ transitions contribute to the QW optical density  $\rho_{\rm qw}(\omega)$ and, in the large-system limit, this becomes a smooth continuous distribution, broadened around $\omega_{\rm qw}$, with a Lorentzian linewidth $\Gamma_{\rm qw}$.
The quality factor of the QW ISB transition is then given by $Q_{\rm qw} = \omega_{\rm qw} /\Gamma_{\rm qw}$.

\begin{figure}
\centering
	\includegraphics[width=\columnwidth]{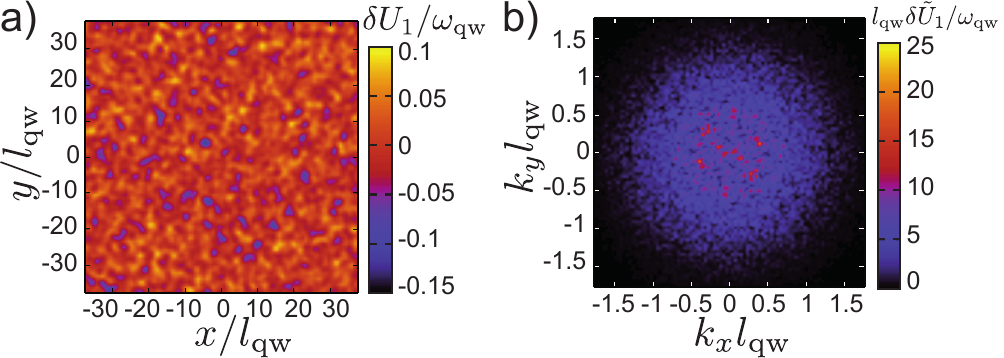}
	\caption{a) Spatial profile of a single realisation of the disorder potential $\delta U_n(\vec{r})$ for the $n=1$ subband. b) Fourier transform of the same disorder potential $\delta \tilde{U}_n(\vec{k})$ as a function of the wavenumber $\vec{k}$.
	Disorder parameters: $\eta_{\rm dis}\approx 0.06$, $\xi_c/l_{\rm qw} =\sqrt{2}$. The numerical methods and the integration parameters are discussed in the Appendix \ref{app:sec:numerical}.
}
	\label{fig:fig2}
\end{figure}

Specifically, we focus here on the case of an infinite well (or square box) confinement in the $z$-direction. In this way we have that
\begin{equation}
\varepsilon_n(L) = \frac{\hbar^2 (\pi n)^2}{2m L^2}.
\end{equation}
We also introduce a reference length scale that depends only from the QW effective electron mass and central frequency
\begin{equation}
    l_{\rm qw} = \sqrt{\frac{\hbar}{ m^* \omega_{\rm qw}}}.
\end{equation}
In the case of an infinite well with length $L_{\rm qw}$, where the central frequency is given by the lowest transition $l_{\rm qw} = \sqrt{2/3} L_{\rm qw}/\pi \approx L_{\rm qw}/4$. From now on we will give all the lengths in units of this length scale.

For simplicity, we specialise on the case of a  Gaussian-distributed interface roughness, for which $\overline{\delta L (\vec{r})} = 0$, and $\overline{ \delta L(\vec{r}) \delta L(\vec{r}^{\, '})} = \xi_0^2/(2\pi) \exp \left[ - |\vec{r} - \vec{r}^{\, '}|^2/\xi_c^2 \right]$.
Here the overline bar indicates the disorder average and $\xi_0$ is the roughness amplitude, while $\xi_c$ is its correlation length.
We can then define a dimensionless parameter which controls the disorder amplitude also in the general case of an arbitrary confinement and an arbitrary disorder roughness
\begin{equation}
    \eta_{\rm dis} =\frac{|\partial_L \varepsilon_1(L_{\rm qw})| \xi_0}{\hbar\omega_{\rm qw}} =  \left(\frac{2}{3} \right)^{3/2}\frac{\xi_0}{\pi l_{\rm qw}} \approx \frac{1}{6}\frac{\xi_0}{l_{\rm qw}},
\end{equation}
where the last equality holds for the specific case of the infinite square box potential (see App. \ref{app:sec:numerical} for major details).
The disorder potential for each subband $n$ reads
\begin{equation}
\delta U_n (\vec{r} )/ \omega_{\rm qw} = n^2  \eta_{\rm dis} \Delta (\vec{r}),
\label{eq:proportional}
\end{equation}
where $\Delta (\vec{r}) = \delta L(\vec{r})/\xi_0$.
In Fig. \ref{fig:fig2}(a) an example of the disorder potential is reported together with its Fourier transform in Fig. \ref{fig:fig2}(b). Here the parameter's choice is inspired by the typical values found in GaAs/AlGaAs systems, see for instance \cite{Unuma_doi:10.1063/1.1535733}.

It is worth to highlight that the use of energy, length units and disorder amplitude in terms of $\hbar \omega_{\rm qw}$, $l_{\rm qw}$ and $\eta_{\rm dis}$ (as detailed in the Appendix \ref{app:sec:numerical}) is important to keep our description as general as possible and to make it independent from the specific shape of the confining QW potential in the direction of the ISB transition ($z$-direction). For instance, the use of these units makes the application to harmonic $z$-confinement case straightforward. The only difference is in how the results are re-parametrised in the disorder $\xi_0$ and correlation $\xi_c$ lengths. Specifically, we will have that $\eta_{\rm dis} = 2\xi_0/l_{\rm qw}$ and $\delta U_n (\vec{r} )/ \omega_{\rm qw} = n\,  \eta_{\rm dis} \Delta (\vec{r})$. Notice the more favourable scaling of the harmonic confinement $\sim n$.

\subsection{Intersubband linewidth at $B=0$}

\begin{figure}
\centering
	\includegraphics[width=\columnwidth]{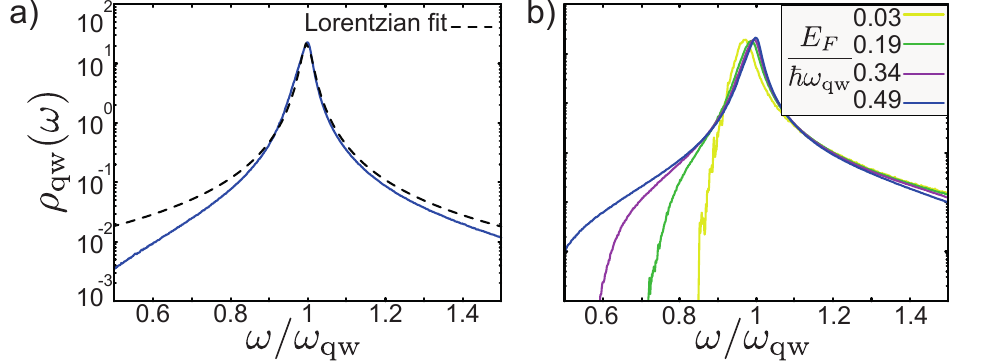}
	\caption{a) Optical density $\rho_{\rm qw}(\omega )$ (blue solid line) and Lorentzian fit (black dashed line). b) Optical density $\rho_{\rm qw}(\omega )$ calculated for increasing values of the Fermi energy $E_F$. 
	Disorder parameters: $\eta_{\rm dis}\approx 0.06$, $\xi_c/l_{\rm qw} = \sqrt{2}$. In a) the Fermi energy is fixed to $E_F/\hbar\omega_{\rm qw} \approx 0.6$, while in b) a range of values is used as indicated in the legend. 
	}
	\label{fig:fig3}
\end{figure}

In the absence of magnetic field, $B=0$, the QW optical density is approximatively given by a Lorentzian distribution \cite{Ando_ISB_space_charge_layers, Bastard_relevance_disorder_ISB_doi:10.1063/1.4766192}, 
\begin{equation}
    \rho_{\rm qw}(\omega ) \approx \frac{1}{2\pi} \frac{\Gamma_{\rm qw}}{\left( \omega - \omega_{\rm qw} \right)^2 + \Gamma_{\rm qw}^2/4},
\end{equation}
where $\Gamma_{\rm qw}$ is its full width at half maximum (FWHM).

This behaviour is confirmed looking at Fig. \ref{fig:fig3}(a), but we also notice some additional features. The long tails of the distribution decay somehow faster than a proper Lorentzian; moreover the optical density $\rho_{\rm qw}(\omega)$ is not completely symmetric and its asymmetry can be increased or reduced by changing the Fermi energy $E_F$, see Fig. \ref{fig:fig3}(b).
In particular reducing the Fermi energy reduces the extension of the low energy tail (on the left side of the peak). All these effects can be understood considering that the left tail is due to the electrons that jump to an energy level of the second subband which has lower energy than their initial level, so to have $\omega^{\parallel}_{k\, k'} < 0$. Assuming the Fermi energy larger than the disorder energy scale, we have that the lowest value of the in-plane electronic transition is approximately given by the Fermi energy $\left[ \omega_{\rm qw} + \omega^{\parallel}_{k\, k'} \right]_{\rm min} \approx \omega_{\rm qw} - E_F/\hbar$. On contrary the value of the Fermi energy has no influence on the tail on the right side of the peak, since the electrons have no limitation in jumping toward higher energies.

\begin{figure}
\centering
	\includegraphics[width=\columnwidth]{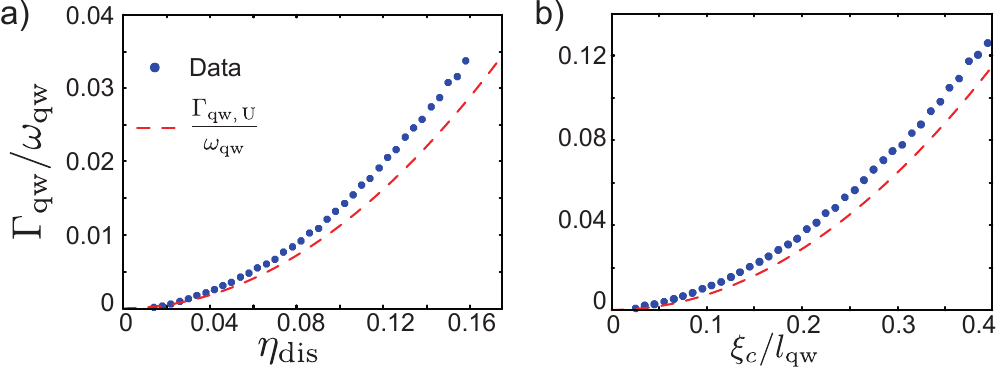}
	\caption{a) Plot of the FWHM $\Gamma_{\rm qw}$ of the ISB optical density $\rho_{\rm qw}$ as a function of the disorder strength $\eta_{\rm dis}$ (blue solid dots).  b) Plot of $\Gamma_{\rm qw}$ as a function of the disorder correlation strength $\xi_c$. Disorder parameters:  $\xi_c/l_{\rm qw} = 0.5$ (in panel a), $\eta_{\rm dis} = 0.4$ (in panel b). Fermi energy $E_F/\hbar \omega_{\rm qw} \approx 0.5$. In both panels, the red dashed line is the Unuma-linewidth given by Eq. \eqref{eq:disorder_strength_estimate_noB}. To numerically implement Eq.\eqref{eq:QW_optical_density} we used a finite-width delta function, with small linewidth $\gamma_{\delta}/\omega_{\rm qw}\approx 0.002$, which is then subtracted from the numerical data.
	}
	\label{fig:fig4}
\end{figure}

As we can see from Fig. \ref{fig:fig4}, the FWHM $\Gamma_{\rm qw}$ scales quadratically both in the disorder strength and in the correlation length \cite{Ando_ISB_space_charge_layers, Unuma_doi:10.1063/1.1535733, Grange_PhysRevApplied.13.044062}
\begin{equation}
\Gamma_{\rm qw}/\omega_{\rm qw} \sim  \eta_{\rm dis}^2\, , \, \xi_c^2/l_{\rm qw}^2,
\end{equation}
as long as the Fermi length is much longer than the disorder correlation length $\xi_c k_F \ll 1$.

Following the semi-analytical approach developed in~\cite{Unuma_doi:10.1063/1.1535733} by Unuma et al., it is possible to derive an analytic expression for the ISB linewidth which holds in the zero energy limit of the electronic scattering against the disorder
\begin{equation}\label{eq:disorder_strength_estimate_noB}
    \frac{\Gamma_{\rm qw,\, U}}{\omega_{\rm qw}} \approx \frac{9}{2} \frac{\xi_c^2}{l_{\rm qw}^2} \eta_{\rm dis}^2.
\end{equation}
As we can see from Fig. \ref{fig:fig4} this formula, that we call \emph{Unuma-linewidth}, fits very well the ISB linewidth extracted from the numerical simulations in the regime of small $\xi_c k_F \ll 1$.

\subsection{Intersubband linewidth at $B\neq 0$}

\begin{figure}
\centering
	\includegraphics[width=\columnwidth]{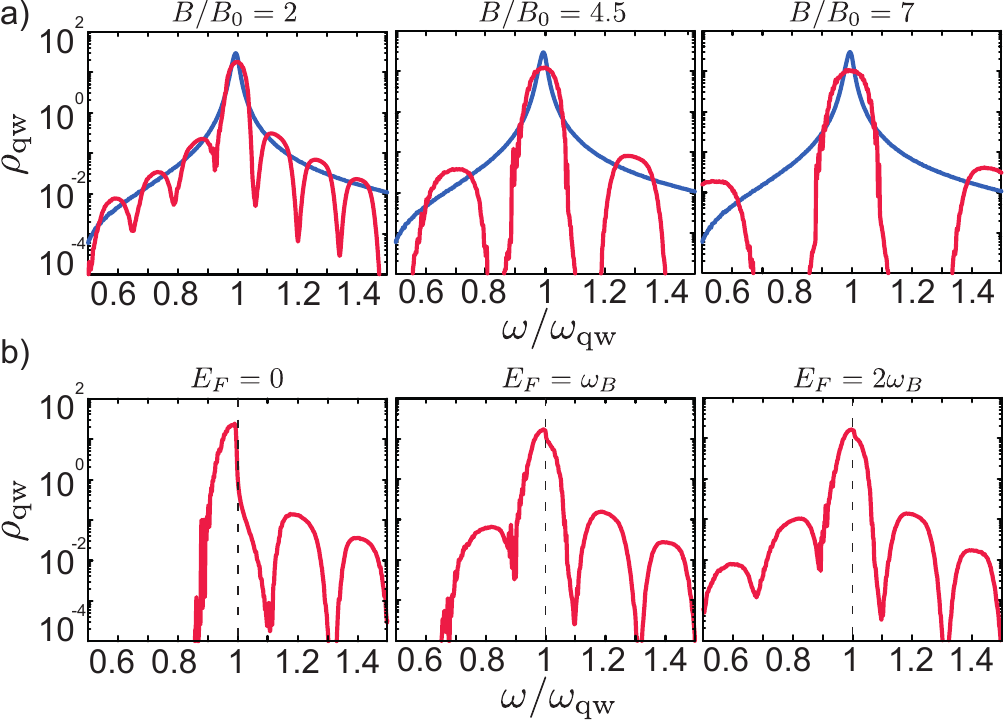}
	\caption{ a) Plots of the optical density $\rho_{\rm qw}(\omega )$ for increasing values of the magnetic field $B/B_0=1,\,4.5,\,7$ (red solid lines). Blue lines show the same quantity in the absence of magnetic field. Disorder parameters: $\eta_{\rm dis}=2/15\approx 0.13$, $\xi_c/l_{\rm qw} = 0.5$. Fermi energy $E_F/\hbar \omega_{\rm qw} = 0.5$. For this choice, the quality factor of the intersubband transition in the absence of magnetic field is numerically found to be $Q_{\rm qw} \approx 40$. b) Plots of the optical density $\rho_{\rm qw}(\omega )$ with fixed magnetic field $B/B_0=2.5$ for increasing values of the Fermi energy $E_F=0, \omega_B, 2\omega_B$. Here the lowest Landau level is set to be at zero energy, such that the Fermi energy falls in the middle of the disorder-broadened $\ell_1 = 0, 1,2$ Landau level. The vertical dashed line highlights the asymmetry of the central Gaussian. Disorder parameters: $\eta_{\rm dis} = 0.08$, $\xi_c/l_{\rm qw} = 1$.  }
	\label{fig:fig5}
\end{figure}
When the magnetic field is turned on, $B \neq 0$, the situation changes drastically. 
In particular when the cyclotron frequency $\omega_B = eB/m^*$ exceeds the energy scale of disorder $\omega_{\rm dis} = \omega_{\rm qw} \eta_{\rm dis} \xi_c/l_{\rm qw}$
we can no longer think about the electrons in terms of free particles diffusing in a disordered landscape, but instead we need to switch to a description in terms of Landau levels.
Having this in mind we can define a reference magnetic field that approximatively sets the border of this transition
\begin{equation}\label{eq:B0}
B_0 = \eta_{\rm dis}\frac{m^* \omega_{\rm qw}}{e}\frac{\xi_c}{l_{\rm qw}} = \eta_{\rm dis}\frac{\xi_c}{l_{\rm qw}} \, \frac{\hbar \omega_{\rm qw}}{2\mu_B^*}
\end{equation}
Here, $\mu_B^*=e\hbar/(2m^*)$ is the effective Bohr magneton relative to the effective mass $m^*$ of the electron.
In the rest of the article we will give all magnetic fields expressed in this rescaled unit. In the last part we will give a more precise analysis on the quantitative conditions to achieve the regime of cavity protection. 

For a completely filled Landau level, 
the Lorentzian optical density is broken into a series of equispaced Gaussian peaks, as we can see from Fig. \ref{fig:fig5}(a). 
The central peak represents all transitions from a given Landau level $\ell_1$ below the Fermi energy in the $1$-th subband, to the same respective Landau level in the $2$nd subband, $\ell_1 = \ell_2$. The side peaks represent instead transitions to other Landau levels of the upper subband, $\ell_2 \neq \ell_1$. For partially filled bands, the peaks corresponding to transitions starting from the uppermost populated Landau level become asymmetric, yet with no qualitative consequence on the overall conclusions of our study. In Fig. \ref{fig:fig5}(b) we plot the optical density with fixed magnetic field when the Fermi energy falls in the middle of the zeroth/first/second broadened Landau level, $E_F = 0, \omega_B, 2\omega_B$. We see that the asymmetry is more prominent when the Fermi energy is in the lowest Landau level, and then is progressively washed out when $E_F$ is in the higher Landau levels.

\begin{figure*}
\centering
	\includegraphics[width=\textwidth]{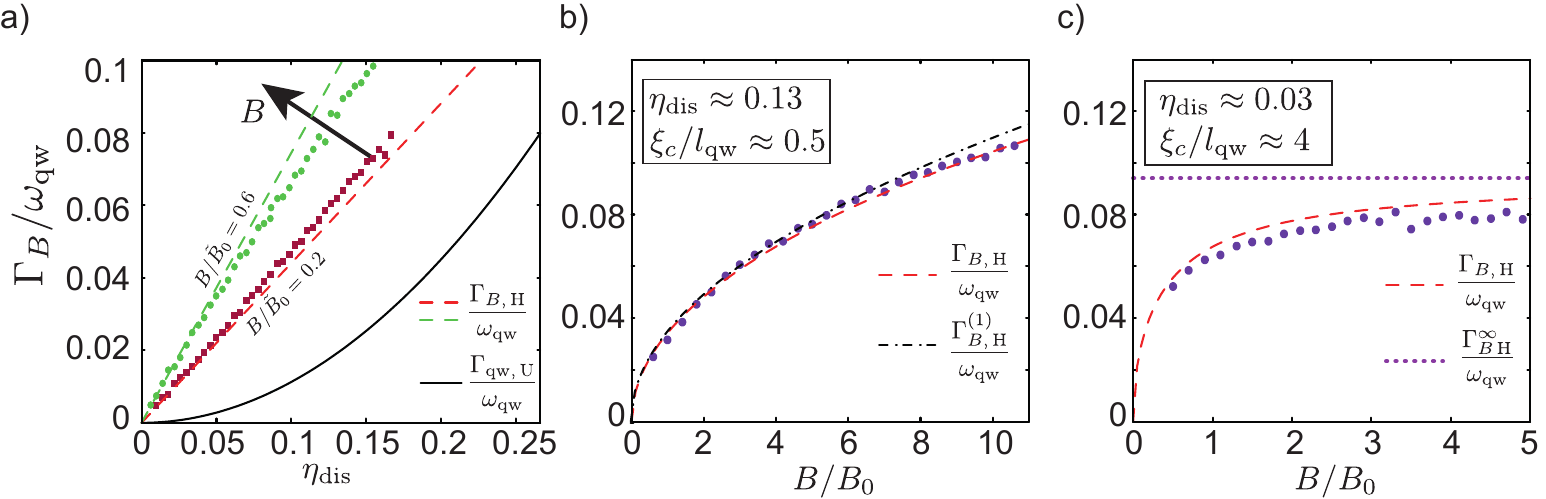}
	\caption{a) Gaussian linewidth (FWHM) $\Gamma_B$ of the central peak of the optical density as a function of the disorder strength $\eta_{\rm dis}$ for a constant value of the magnetic field $B$. The red squares/green dots are the numerical data corresponding to increasing values of $B$, that is $B/\tilde{B}_0 = 0.2$, $B/\tilde{B}_0 = 0.6$. The disorderless reference magnetic field is defined as $\tilde{B}_0 = \hbar \omega_{\rm qw}/(2\mu_B^*)$. 
	The red/green dashed lines are given by the Hikami linewidth $\Gamma_{B, \, {\rm H}}/\omega_{\rm qw}$ while the black solid line is the Unuma linewidth $\Gamma_{\rm qw, \, U}/\omega_{\rm qw}$. Disorder parameters: $\xi_c/l_{\rm qw} = 0.5$.  b) Gaussian linewidth $\Gamma_B$ (purple dots) extracted from numerical simulations of the optical density as a function of the magnetic field $B$. The dashed red line is given by the Hikami linewidth $\Gamma_{B, \, {\rm H}}/\omega_{\rm qw}$, while the dashed-dotted line is given by the small disorder expansion of the Hikami linewidth $\Gamma_{B, \, {\rm H}}^{(1)}$. Disorder parameters as in Fig. \ref{fig:fig5}: $\eta_{\rm dis} = 2/15\approx 0.13$ and $\xi_c/l_{\rm qw} = 0.5$.
	c) Same plot as in b) but with different parameters in order to highlight the saturation of the linewidth at high magnetic field. The red dashed line is given by the Hikami-linewidth $\Gamma_{B, \, {\rm H}}/\omega_{\rm qw}$. The purple dotted line is the saturation value of the Hikami linewidth given by $\Gamma_{B, \, {\rm H}}^{\infty}/\omega_{\rm qw} \approx 0.094$. Disorder parameters: $\eta_{\rm dis} = 1/30 \approx 0.03 $ and $\xi_c/l_{\rm qw} = 4$.
	In all plots the Fermi energy is $E_F = 1.5 \hbar \omega_B$, in such a way that the central Gaussian is determined by only the two lowest Landau levels.
	To numerically implement Eq.\eqref{eq:QW_optical_density} we used a finite-width delta function, with small linewidth $\gamma_{\delta}/\omega_{\rm qw}\approx 0.003$, which is subtracted from the numerical data.}
	\label{fig:fig6}
\end{figure*}

The spectral distance between neighbouring Gaussian is approximately given by the cyclotron frequency, $\sim \omega_B$, as a consequence of the transitions between neighbouring Landau levels.
Focusing on the properties of the central Gaussian, centered on $\omega_{\rm qw}$, we call its linewidth $\Gamma_B$.
Contrary to the non magnetic case, discussed in the previous section, here we expect that the linewidth of each of these Gaussian lobes scales linearly with the disorder strength
$\Gamma_{B}/\omega_{\rm qw} \sim  \eta_{\rm dis}$.
This can be understood from the fact that the disorder plays the role of a small perturbation on a degenerate system, bringing corrections at the linear order instead of the usual second order.

As in the non-magnetic case, the disorder correlation length $\xi_c$ plays an important role in determining the width of each Gaussian, but in contrast to the non-magnetic case, here the system has an intrinsic length scale, given by the magnetic length
\begin{equation}
l_B = \sqrt{\frac{\hbar}{e B}}.
\end{equation}
We then expect that the FWHM of each Gaussian depends non-trivially on $\xi_c/l_B$.

Our aim is now to have more quantitative insights on the dependence between the width of the Gaussian lobes and the system parameters.
In order to do so we start from the observation that not all the transitions are important in this regime of strong magnetic field.
Indeed, considering the central Gaussian lobe of the optical density, we realise that only the intra-Landau-level transitions with the same $k_n$ are actually relevant.
This is justified by the fact that the disorder landscape is the same in the two subbands, and differs only in its amplitude. Specifically, in the $2$-subband it is $4$ times larger than in the $1$-subband.  
The system is then strongly localised but each state in the ground subband can overlap only with the state in the upper subband localised in the same region. 
So, in Eq. \eqref{eq:QW_optical_density}, we can approximate $\Lambda_{(\ell_1 k_1), (\ell_2, k_2)}\approx \delta_{\ell_1 \ell_2}\delta_{k_1 k_2}$. The energy difference of each transition from a state $(\ell_1, k_1)$ in the $1$-subband to the corresponding state $(\ell_2,k_2)$ in the $2$-subband is then given by $\omega_{\rm qw} + \omega_{(\ell_2 k_2) (\ell_1 k_1)}^{\parallel} \approx \omega_{\rm qw} + 3 (\omega_{\ell_1 k_1} - \omega_B\ell_1)$, where $\omega_{\ell_1 k_1}$ is the $k_1$-th eigenenergy of the $\ell_1$-th Landau level of the $1$-subband.

The considerations above are particularly important when we restrict each band to only include the lowest Landau level (LLL). This is well justified when the Fermi energy is smaller than the cyclotron frequency, $E_F < \hbar \omega_B$, and it is a crucial assumption in order to carry on the calculation analytically. However, we will see that the result extracted in the LLL remains a very good approximation also in the more general case.
Calling the linewidth of the LLL $\Gamma_{LLL}$ we arrive to conclude that the linewidth of the optical density is approximately three times the linewidth of the $1$-subband LLL, $\Gamma_B \approx 3\, \Gamma_{LLL}$. 
In \cite{hikami_1985_:jpa-00210150} we can find an exact expression for $\Gamma_{\rm LLL}$ extracted from the density of states of a non-interacting 2-dimensional electron gas in presence of a Gaussian-correlated random potential in the limit of large correlation length $\xi_c/l_B \gg 1$ (a very similar result can be derived also for the case of Gaussianly distributed short-range scatters, see \cite{Ando_ISB_space_charge_layers, Ando_broadening_doi:10.1143/JPSJ.39.279,Ando_localization_doi:10.1143/JPSJ.52.1740}). 
A brief summary of the calculation is reported in the Appendix \ref{app:sec:disordered_densityLLL}.
Since the resulting density of state is Gaussian, the FWHM reads
\begin{equation}\label{eq:LLL_FWHM}
    \frac{\Gamma_{\rm LLL}}{\omega_{\rm qw}} = 2\sqrt{\frac{\log (2)}{\pi} }\, \frac{\xi_c/l_B}{\sqrt{\xi_c^2/l_B^2 + 2}} \, \eta_{\rm dis}.
\end{equation}
The FWHM of the central Gaussian peak of the ISB optical density is thus given by
\begin{equation}\label{eq:B_linewidth_approx_scaling}
\frac{\Gamma_{B, \, {\rm H}}}{\omega_{\rm qw}} \approx 6\sqrt{\frac{\log (2)}{\pi} } \,  \frac{\xi_c/l_B}{\sqrt{\xi_c^2/l_B^2 + 2}} \, \eta_{\rm dis}.
\end{equation}
This formula, that we call \emph{Hikami-linewidth}, fits extremely well the optical density's FWHM extracted numerically as we can see from Fig. \ref{fig:fig6}, even in the case in which higher Landau level contribute to the optical density.

As expected the width of the central Gaussian scales linearly with the disorder strength $\sim \eta_{\rm dis}$, which is is perfectly captured from Eq. \eqref{eq:B_linewidth_approx_scaling}, see Fig. \ref{fig:fig6}(a). 
In this plot we also included the corresponding values from the Unuma linewdith $\Gamma_{\rm qw, \, U}$ from Eq. \eqref{eq:disorder_strength_estimate_noB}, which gives an estimate of the ISB linewidth without the magnetic field. 
It is worth noticing that at this stage the effect of the magnetic field is actually to broaden the ISB transitions and so to worsen the quality factor of our bare QW. In the next section we will see that an opposite result occurs when the QW is embedded in a cavity. 

For small values of the magnetic field we also found a linear scaling for the width in $\sim\xi_c/l_B$, which means a square-root scaling in the magnetic field intensity $\sim \sqrt{B}$, as can be seen from Fig. \ref{fig:fig6}(b).
Quite surprisingly we realise that the Hikami formula gives an accurate estimation of the linewidth even in this regime of moderate correlation length, where $\xi_c \lesssim l_B$. Expanding it at lowest order in $\xi_c/l_B$ we recover the square-root behaviour in the magnetic field strength. Using Eq. \eqref{eq:disorder_strength_estimate_noB} we can re-express the small disorder expansion of the Hikami linewidth in a nice and compact expression
\begin{equation}\label{eq:Gamma_B_sqrt_approx}
    \frac{\Gamma_{B,\,{\rm H}}}{\omega_{\rm qw}} \approx \frac{\Gamma_{B,\,{\rm H}}^{(1)}}{\omega_{\rm qw}} = 2\sqrt{\frac{\log (2)}{\pi} }\,  \sqrt{\frac{2\mu_B^* B}{\hbar \omega_{\rm qw}}} \sqrt{\frac{\Gamma_{\rm qw\, U}}{\omega_{\rm qw}}} .
\end{equation}
In Fig. \ref{fig:fig6}(b) we see that both Eqs. \eqref{eq:B_linewidth_approx_scaling}-\eqref{eq:Gamma_B_sqrt_approx} fits very well the numerical data in our regime of interest.
It is worth noticing once again the broadening effect of the magnetic field, by comparing the Hikami linewidth to the Unuma linewidth $\Gamma_{B, \, {\rm H}}^{(1)}/\Gamma_{\rm qw, \, U} = 2\sqrt{\log (2)/\pi} \sqrt{\omega_B/\omega_{\rm qw}}\sqrt{Q_{\rm qw}}$. Even if the typical magnetic field that we are using through this work is such that $\omega_B/\omega_{\rm qw}\sim 0.1 - 0.5$, the ISB quality factor is always taken to be $Q_{\rm qw} \gtrsim 10$, in order to fulfil the condition of small disorder, and so $\Gamma_{B, \, {\rm H}}^{(1)}/\Gamma_{\rm qw, \, U} \gtrsim 1$. 

In Fig. \ref{fig:fig6}(c) instead the behaviour at larger values of $\xi_c/l_B$ is reported, and we see how the linewidth saturates for large magnetic field, approaching the asymptotic value
\begin{equation}\label{eq:hikami_LW_infinite}
    \frac{\Gamma_{B \, {\rm H}}^{\infty} }{\omega_{\rm qw}} \approx 6\sqrt{\frac{\log (2)}{\pi} }\, \eta_{\rm dis}.
\end{equation}

\begin{figure}
\centering
	\includegraphics[width=\columnwidth]{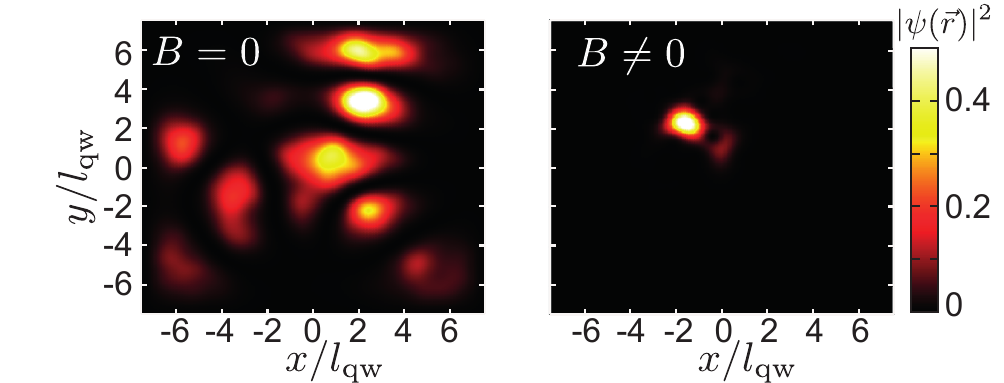}
	\caption{Illustrative examples of the real-space wavefunction of a $n=1$ subband eigenstate in the presence of disorder in the $B=0$ (left) and $B/B_0 = 0.5$ (right) cases. These eigenstates are obtained via exact diagonalisation, as detailed in App. \ref{app:sec:numerical}, of a square system with lateral size $L_{x,y}/l_{\rm qw} = 14$. Other parameters: $E_F /(\hbar \omega_{\rm qw}) = 0.5$, $\eta_{\rm dis} \approx 0.03$, $\xi_c/l_{\rm qw} \approx 4 $.}
	\label{fig:fig7}
\end{figure}

\begin{figure*}[t]
\centering
	\includegraphics[width=\textwidth]{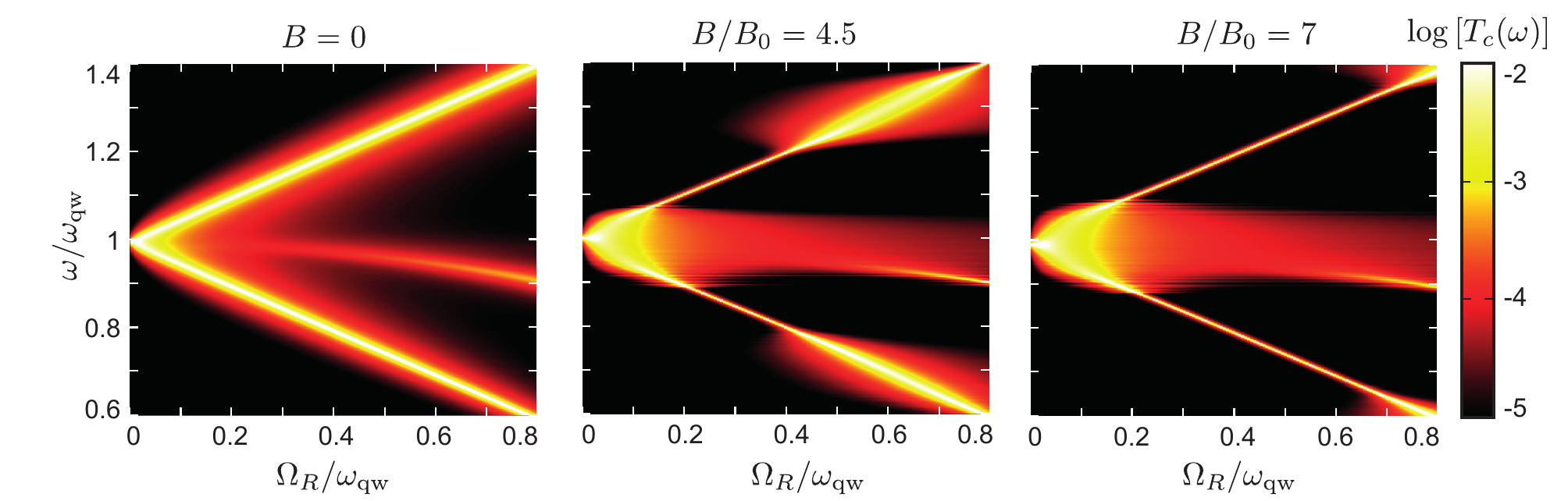}
	\caption{ Logarithmic colorplot of the cavity transmission as a function of the Rabi coupling $\Omega_R$ and the external incident frequency $\omega$ for various values of the magnetic field $B$ in the resonant regime $\omega_c = \omega_{\rm qw}$. 
Disorder parameters as in Fig. \ref{fig:fig5}: $\eta_{\rm dis} = 2/15 \approx 0.13$, $\xi_c/l_{\rm qw} = 0.5$. Fermi energy $E_F /\hbar \omega_{\rm qw}= 0.5$.The ISB quality factor (without cavity) in the three panels is numerically found to be $Q_{\rm qw}\approx 40, 12, 10$, from left to right.
An extreme value of the cavity quality factor $Q_c = 10^5$ is taken for illustrative purposes, more realistic values will be considered in the next figures.
In physical units, for a QW of thickness $L_z = 26\,$ nm and $m^* \approx 0.067\, m_e$, these adimensional parameters correspond to $\hbar \omega_{\rm qw} \approx 25\,$meV and a reference magnetic field $B_0 \approx 1 \,$ T. In the central panel we thus have $B \approx 4.4\,$T while in the right panel $B \approx 6.9\,$T.  }
	\label{fig:fig8}
\end{figure*}

\subsection{Discussion}

As we have seen in the previous discussion, there is a large difference in the optical properties of the ISB QW between the non-magnetic and the strong magnetic case. In particular the magnetic field turns the shape of the optical density from a Lorentzian to a Gaussian (a central Gaussian separated from a series of smaller Gaussian side lobes), and sensibly broadens the ISB transition.
These differences can be traced back to the localisation properties of the electrons in the disordered potential due to the interface roughness. 

Without the magnetic field the electronic states are partially localised due to the interplay of multiple scattering and interference processes on the complex landscape of the disorder potential. 
The Lorentzian shape of the optical density can be then understood in terms of the finite effective lifetime that the scattering on disorder gives to the electronic coherence. 

When the magnetic field is turned on the kinetic energy is completely quenched and we switch to a new regime of strong spatial localisation in Landau levels, where the localisation length is set by the magnetic length $l_B$. 
In Fig. \ref{fig:fig7} we can see an example how eigenfunctions at similar eigenenergies localize in a very different way in the two cases. 
An alternative, possibly more intuitive understanding of this physics can be obtained within a semiclassical picture: in the strong magnetic field regime with $l_B \ll \xi_c$, the electrons follow the semi-classical guiding center trajectories along the equipotential lines of the disorder of each subband\cite{Huckestein_1995_RevModPhys.67.357}. Their energy levels are approximately given by $\hbar \omega_{\ell \, k}\sim \hbar \omega_B \ell + \delta U(\vec{r}_k)$, where for each value of the quantum number $k$ labelling the states, one has to pick a different representative real-space position $\vec{r}_k$ along the orbit.
Since according to \eqref{eq:proportional} the disorder potential felt by the two subbands are proportional to each other, the electronic transitions are only between states with the same localization pattern (as described in the last section) and their frequencies are effectively sampling the values of the disorder potential, which follows a Gaussian distribution.
This gives a basic intuitive explanation why the resulting optical density is Gaussian-distributed with a linewidth set by the width of the disorder potential. 
Despite the much larger linewidth, it is the much faster decay of the tails of the Gaussian distribution compared to the ones of a Lorentzian distribution which will be at the heart of our developments in the next sections.

\section{Intersubband polaritons and cavity protection}
\label{sec:Intersubband polaritons and cavity protection}
In this section we show how the dramatic change in the shape of the QW optical density $\rho_{\rm qw}(\omega )$ due to the magnetic field that was displayed in Fig. \ref{fig:fig5} can be used to sensibly improve the properties of the ISB cavity polaritons.
It is in fact a well-known fact that a sufficiently strong coupling of a collection of emitters with an overall Gaussian optical density to a single-mode cavity may give rise to polaritonic peaks whose linewidth is only limited by the cavity losses~ \cite{FirstCavityProtection_Houdre_PhysRevA.53.2711}. On contrary, the linewidth of the polariton peaks resulting from an emitter with a Lorentzian-shaped optical density is set by the average of the cavity and emitter linewidths. The physical mechanism underlying these two different behaviours goes under the name of {\em cavity protection}~\cite{ Diniz_CavityProtection_PhysRevA.84.063810, Molmer_CavityProtection_PhysRevA.83.053852}, and is a general feature of polaritonic systems independently of their material realization~\cite{Mayer_CavityProtection_cQED_exp_NaturePhys, Faraon_cavity_protection_exp_nanoph_cavity, Breeze_cavity_protection_molecules, Colombelli_immunity_PhysRevB.96.235301, Long_arxiv_cavityProtection2022_https://doi.org/10.48550/arxiv.2208.12088}.

\subsection{Magnetic-field-induced cavity protection}

We consider here the case in which the magnetic field is strong and exceeds the disorder strength, $B > B_0$.
Since in a clean sample the polariton frequencies are given by $\omega_{\pm} = \omega_c \pm \Omega_R/2$,
when the Rabi frequency becomes comparable with the cyclotron frequency $\Omega_R \sim \omega_B$, the polariton modes sit in between a pair of Gaussian peaks of the optical density. Here, the optical density has a much smaller value than in the standard non-magnetic case, $\rho_{\rm qw}(\omega_{\pm})|_{B\neq 0}\ll \rho_{\rm qw}(\omega_{\pm})|_{B= 0}$. 
If we think of the system in terms of the clean polariton eigenstates coupled to another continuum of states given by the disorder, we can apply the standard Weisskopf-Wigner theory, expecting that the polaritonic linewidth $\Gamma_{\pm}$ is proportional to the optical density calculated at the polariton frequency, $\Gamma_{\pm} \sim \rho_{\rm qw}(\omega_{\pm})$.
If the optical density drops in correspondence of the polariton frequencies a strong linewidth-narrowing effect is expected. 

This behaviour is indeed clearly visible in the numerical results shown in Fig. \ref{fig:fig8}, where we plot the cavity transmission $T_c(\omega)$ defined in Eq. \eqref{eq:cavity_transmission} as a function of the transmitted frequency $\omega$ and the Rabi frequency $\Omega_R$. It is clear that, for a sufficiently large value of $B/B_0$ and for a Rabi frequency comparable to the cyclotron frequency $\Omega_R \approx \omega_B$, the polaritonic transmission linewidth is several order of magnitude smaller than the $B=0$ case and is only limited by the cavity linewidth.

\begin{figure}
\centering
	\includegraphics[width=\columnwidth]{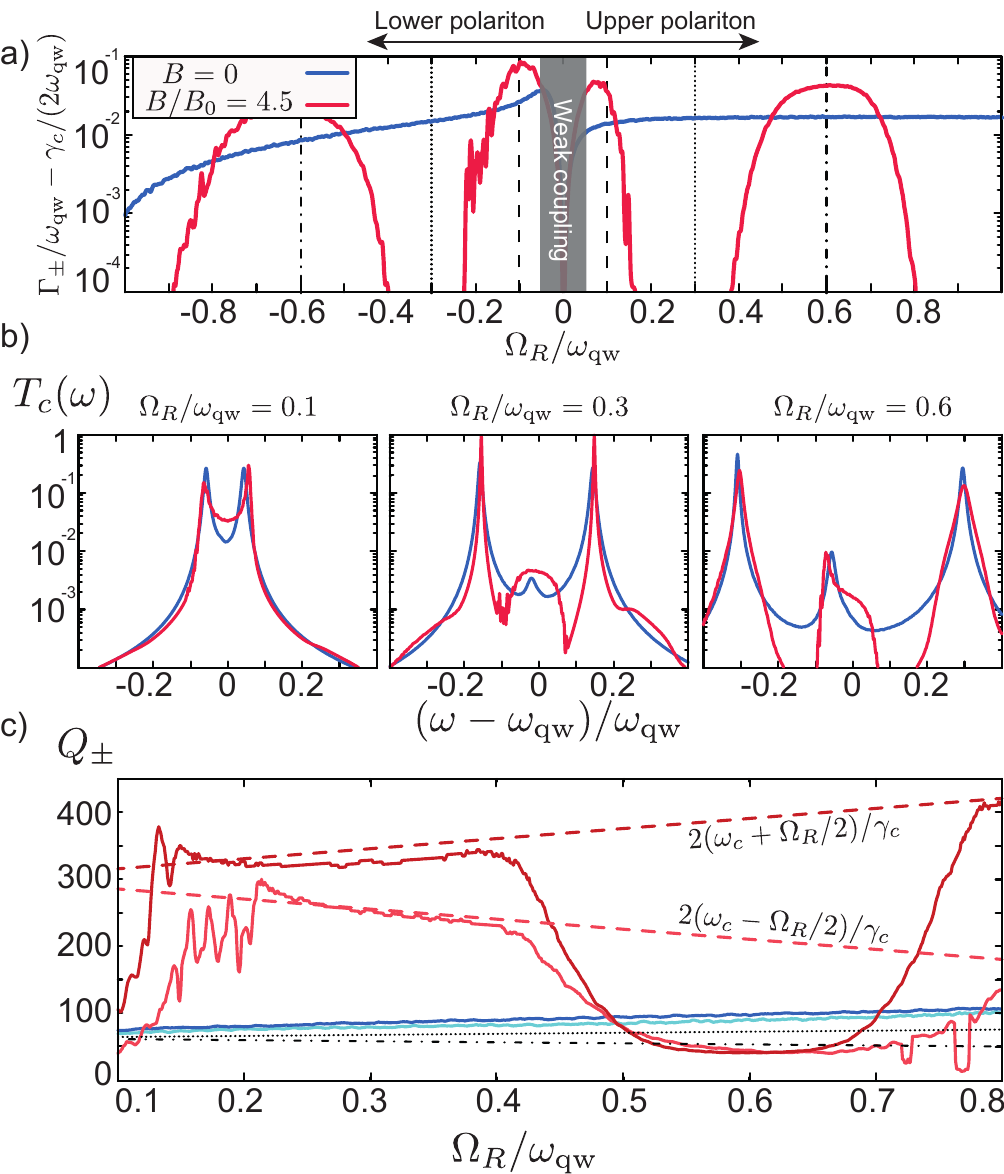}
	\caption{ a) Plot of the electronic contribution to the polariton linewidth estimated using Eq. \eqref{eq:linewidth_polariton_diniz} as a function of Rabi coupling $\Omega_R$, left (right) parts corresponding to the upper (lower) polariton. 
The blue and red lines refer to the $B=0$ and $B=4.5\, B_0$ cases, respectively. b) Three examples of the cavity transmission $T_c(\omega)$ given by Eq. \eqref{eq:cavity_transmission} as a function of the incident frequency $\omega$ in the presence/absence of magnetic field (blue and red lines, respectively) for increasing values of the Rabi coupling $\Omega_R$ (from left to right) as indicated by he dashed, dotted and dashed-dotted lines in panel a). 
	c) The quality factor of the lower (light) and upper (dark) polaritons $Q_{\pm} = \omega_{\pm}/\Gamma_{\pm}$ as a function of the Rabi frequency $\Omega_R$. Blue and red lines refer to the cases in the absence/presence of magnetic field $B/B_0 = 0, 4.5$. The black dotted and black dashed-dotted lines indicate the usual upper/lower polariton quality factors obtained averaging between cavity and ISB linewidths, $Q_{\pm, \, {\rm std}} = 2(\omega_{c} \pm \Omega_R/2)/(\gamma_c + \Gamma_{\rm qw})$. The light/dark red dashed lines indicate the upper bound to the polariton quality factor set by the cavity losses as defined in Eq. \eqref{eq:Qmax_pol}. Same system parameters as in Fig.\ref{fig:fig8} except for the realistic value $Q_{c} = 150$ of the cavity quality factor. 
 }
	\label{fig:fig9}
\end{figure}

A further increase of $\Omega_R$ makes the polaritonic transmission to broaden again as in the weak Rabi frequency case. This happens specifically when the polariton frequencies correspond to one of the side Gaussians of the optical density, that is for $\Omega_R \sim 2n\cdot \omega_B$, with $n=1,2\ldots \,$.
This behaviour is well captured following the theory developed by Diniz et al. \cite{Diniz_CavityProtection_PhysRevA.84.063810}, where the polaritonic linewidth can be estimated using the following formula
\begin{equation}\label{eq:linewidth_polariton_diniz}
    \Gamma_{\pm} \approx \frac{1}{2}\left[{\gamma_c + \frac{\pi}{2}\Omega_R^2 \rho_{\rm qw} \left( \omega_c \pm \frac{\Omega_R}{2}\right)}\right],
\end{equation}
where $\gamma_c$ is the cavity linewidth.
This formula holds in the strong coupling regime, when $\Omega_R \gg \gamma_c, \Gamma_{\rm qw}$ (see App. \ref{app:pol_linewidth} for more details).
Having $\Omega_R > \gamma_c, \Gamma_{\rm qw}$ means that the polaritonic linewidth $\Gamma_{\pm}$ is determined by the shape of the tails of the QW optical density. If $\rho_{\rm qw}$ has a Lorentzian shape, as in the regular non-magnetic case, we have that $\pi/2\, \Omega_R^2 \rho_{\rm qw} \left( \omega_c \pm \Omega_R/2\right) \approx \Gamma_{\rm qw}$ and the resulting polaritonic linewidth is the average between the cavity and ISB linewidths, $\Gamma_{\pm } \approx (\gamma_c + \Gamma_{\rm qw})/2$.
On contrary, if the QW optical density has a Gaussian shape (or, more generally, decays faster than $1/\omega^2$), we have that $\pi/2\, \Omega_R^2 \rho_{\rm qw} \left( \omega_c \pm \Omega_R/2\right) \approx 0$ and the resulting polaritonic linewidth is only given by the contribution due to the cavity linewidth $\Gamma_{\pm} \approx \gamma_c/2$.

In Fig. \ref{fig:fig9}(a) we plot the result of \eqref{eq:linewidth_polariton_diniz} in the two cases of $B=0$ and $B\approx 4.5\, B_0$. When the vacuum Rabi coupling $\Omega_R$ is comparable to the QW linewidth $\Gamma_{\rm qw}$ the polaritonic peaks have similar linewidth in both the non-magnetic and magnetic cases, Fig. \ref{fig:fig9}(b) left panel. Instead, when the strong coupling regime is fully reached the mechanism behind the cavity protection sets in: the contribution of the ISB transition to the polariton linewidth drops to zero, as shown in Fig \ref{fig:fig9}(a), and the linewidth is only given by half the cavity linewidth. In Fig. \ref{fig:fig9}(b) central panel, we illustrate the cavity protection effect for a cavity quality factor $Q_{c} \approx 150$ close to the typical experimental values. However, as pointed out at the beginning of the section, this is true only when the polariton frequency is located in a gap between two Gaussian peaks, i.e. when $\Omega_R \sim \omega_B$: when the polariton frequency reaches the next Gaussian peak, at $\Omega_R=2\omega_B$ the linewidth is again broadened to its non protected value, see the right panel of Fig. \ref{fig:fig9}(b).

In Fig. \ref{fig:fig9}(c) we plot the quality factors of the lower and upper polaritons  defined as
\begin{equation}
    Q_{\pm} = \frac{\omega_{\pm}}{\Gamma_{\pm}},
\end{equation}
as a function of the Rabi frequency, in the presence and in the absence of the magnetic field.
As in the other plots of Fig \ref{fig:fig9}, we set $\omega_c = \omega_{\rm qw}$, we take the cavity quality factor as $Q_c = \omega_c/\gamma_c = 150$ and, when it is on, the magnetic field is set to $B/B_0 = 4.5$.
We extract numerically the polaritonic frequencies $\omega_{\pm}$ and the linewidths $\Gamma_{\pm}$ as, respectively, the peak frequencies of the cavity transmission $T_c(\omega )$, and the FWHM of each peak.
When the magnetic field is on, the polaritonic quality factor increases sensibly (the light and dark solid lines in Fig. \ref{fig:fig9}(c)), reaching its limiting value
\begin{equation}\label{eq:Qmax_pol}
    Q_{\pm\,{\rm max}} = 2\frac{\omega_c \pm \Omega_R/2}{\gamma_c},
\end{equation}
marked by the light and dark red dashed lines. For instance, when $\Omega_R/\omega_{\rm qw} \approx 0.3$ we have that the polaritonic quality factors increase by a factor between $3$ and $4$. Of course, the improvement factor would be far more dramatic for higher values of the cavity quality factors $Q_c$ as shown in Fig.\ref{fig:fig8}. Such improvements are presently the subject of intense research~\cite{Abujetas:19, Pisani:20, Chalimah_PhysRevApplied.15.064076, van_Hoof_BICTHz_2022}.

Moreover, we have to notice that even in the absence of magnetic field both lower and upper polaritonic quality factors $Q_{\pm}$ might display a slight increase as a function of the Rabi frequency $\Omega_R$ (light and dark blue solid lines in Fig. \ref{fig:fig9}(c)). This is a side effect due to the non-perfectly Lorentzian and asymmetric shape of the QW optical density $\rho_{\rm qw}(\omega )$. However this is a rather marginal effect if compared to the dramatic linewidth suppression that is induced by the magnetic field.

\begin{figure}
    \centering
    \includegraphics[width=\columnwidth]{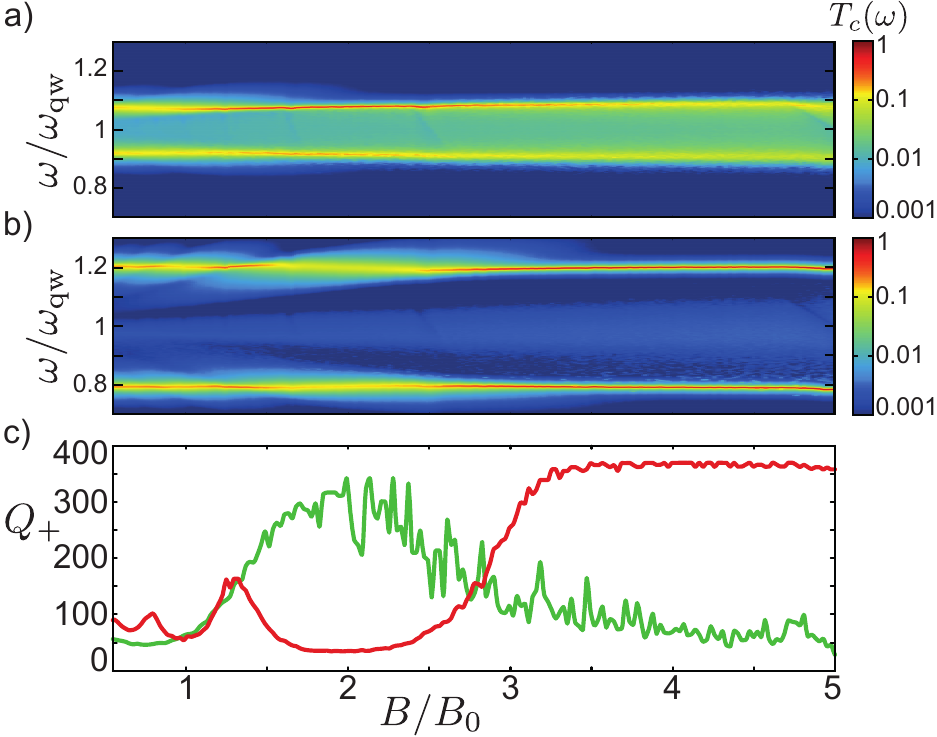}
    \caption{a-b) Logscale color plot of the cavity transmission $T_c(\omega)$ as a function of the magnetic field $B/B_0$ and the external incident frequency $\omega/\omega_{\rm qw}$. In a) $\Omega_R/\omega_{\rm qw} = 0.15$, in b) $\Omega_R/\omega_{\rm qw} = 0.4$. c) Quality factor of the upper polariton as a function of the magnetic field $B/B_0$. The green and red solid lines refer to the $\Omega/\omega_{\rm qw} = 0.15$ and $0.4$ values used in panel a) and b), respectively. Parameters for all plots: $\omega_c = \omega_{\rm qw}$, $\eta_{\rm dis} = 0.1$, $\xi_c/l_{\rm qw} = 1$, Fermi energy $E_F/\hbar \omega_{\rm qw} = 0.5$. For these parameters, the ISB quality factor (without cavity) is found to be $Q_{\rm qw} \approx 22$, while the cavity quality factor is set to the realistic value $Q_c = 200$.}
    \label{fig:fig9plus}
\end{figure}

As a final point we investigate the behaviour of the cavity transmission $T_c(\omega )$ as a function of the external magnetic field $B$, keeping the Rabi frequency $\Omega_R$ constant. 
In Fig. \ref{fig:fig9plus}(a-b) we  respectively choose $\Omega_R/\omega_{\rm qw} = 0.15$ and $\Omega_R/\omega_{\rm qw} = 0.4$ and we sweep the magnetic field in the range $B/B_0 \approx 0.5 - 5$. 
In order to have a better readability of the figure, the magnetic-field-dependence of the quality factors of the upper polariton in the two cases is summarized in the bottom panel Fig. \ref{fig:fig9plus}(c) -- the quality factors of the lower polaritons follow a very similar trend, so, for clarity, are not reported. 
When the Rabi frequency is small, e.g.  the $\Omega_R/\omega_{\rm qw} = 0.15$ value used in Fig. \ref{fig:fig9plus}(a), the magnetic-field-induced cavity protection competes against the magnetic linewidth broadening of the ISB transition, calculated in Eq. \eqref{eq:B_linewidth_approx_scaling}. As a result, the polaritonic quality factor $Q_+$ increases with the magnetic field to a maximum and then decreases again to its minimum value, see the green solid line in Fig. \ref{fig:fig9plus}(c).
On the other hand, for a larger value of the Rabi frequency, e.g. the $\Omega_R/\omega_{\rm qw} = 0.4$ value used in Fig. \ref{fig:fig9plus}(b), the polaritonic quality factor $Q_+$ saturates to its maximum value, forming a plateau that extends for a rather wide range of magnetic fields, see the red solid line in Fig. \ref{fig:fig9plus}(c). We also notice that in this regime of large Rabi frequency, additional oscillations are visible in the polaritonic quality factor for smaller values of magnetic fields. Here, the maxima correspond to values of the Rabi frequency matching the gap between two higher Landau levels.

\subsection{Experimental implementations}

\begin{figure}
    \centering
    \includegraphics[width=\columnwidth]{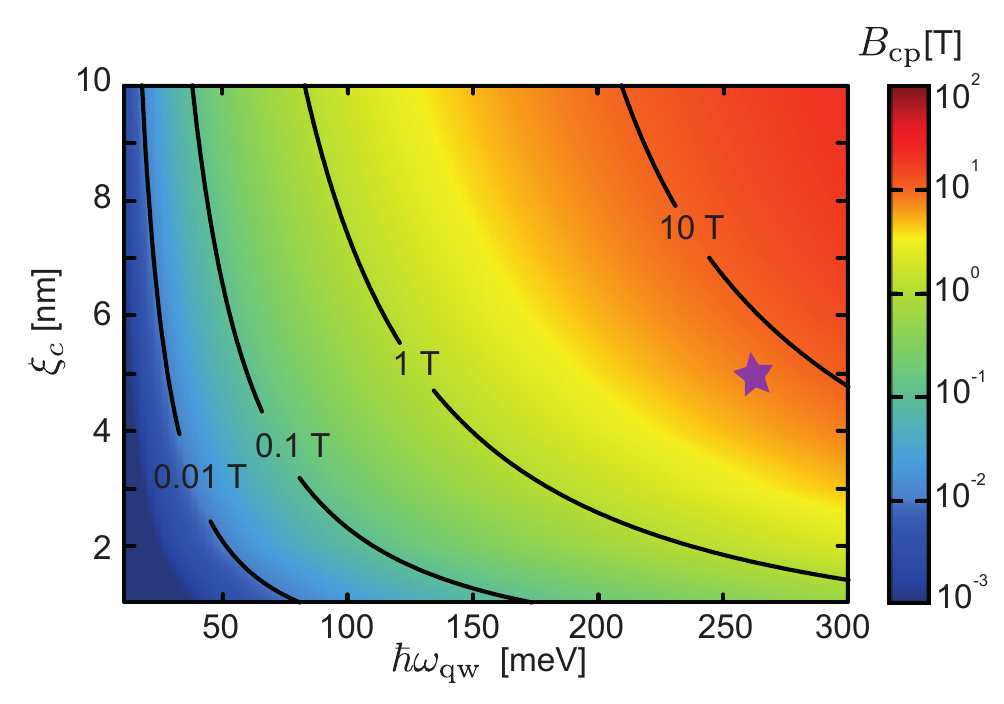}
    \caption{Log-scale contour-plot of the cavity-protection magnetic field $B_{\rm cp}$ as defined in \eqref{eq:Bcp} as a function of the intersubband frequency $\omega_{\rm qw}$ and the disorder correlation length $\xi_c$. Here, $B_{\rm cp}$ is given in units of Tesla, $\hbar\omega_{\rm qw}$ is given in meV, and $\xi_c$ in nm. The disorder amplitude is kept constant to $\xi_0 = 0.3 \,$nm. The effective Bohr magneton corresponding to the effective mass of electrons in GaAs is used, $\mu_B^* \approx 0.86 \,$meV / T. The purple star indicates the position of the setup inspired by \cite{Unuma_doi:10.1063/1.1535733} described in the text. }
    \label{fig:fig10}
\end{figure}

In this section we discuss the actual experimental feasibility of the magnetically induced cavity protection for ISB polaritons. In doing this, we will focus on the minimal values of the magnetic field and of the Rabi frequency that are needed to observe the effect.
We restrict to devices operating in the THz or mid-infrared regime, mainly focusing on the parameters reported in \cite{Colombelli_Perspectives_PhysRevX.5.011031}.
A typical size for a QW in this regime is $L_z \sim 8-40\,$nm. Using the GaAs electron's effective mass $m^* \approx 0.067\, m_e$, where $m_e$ is the free electron mass, we have that the energy range for the fundamental ISB transition is given by
\begin{equation}
    \hbar \omega_{\rm qw} \sim 10-250\, {\rm meV}.
\end{equation}
The typical roughness fluctuation scale is around $\xi_0 \sim 0.1 - 1\,$nm and its typical correlation length is $\xi_c \sim 1-10 \,$nm. 
The typical quality factor of the bare ISB transitions in these devices is in the range $Q_{\rm qw} \sim 10-50$. 


In order to estimate the typical magnetic field and Rabi frequency needed to quench the linewidth, there are two conditions that we need to satisfy: 
\begin{enumerate}
    \item We need to break the Lorentzian optical density in well separated Gaussians. This is achieved when 
    \begin{equation}
        \omega_B > \Gamma_B.
    \end{equation}
    \item We need to have a Rabi frequency that is large enough to overcome the linewidth of the central Gaussian of the optical density and, thus,  reach the strong-coupling regime. This condition is maximally fulfilled when the polariton frequency falls in the middle of a gap between two Landau levels. This occurs when 
    \begin{equation}\label{eq:OmegaRcp}
        \Omega_R = \omega_B.
    \end{equation}
\end{enumerate}
Because of the very fast decay of the Gaussian tails of the magnetic optical density, the first condition can be considered fulfilled already when $\omega_B = 2\, \Gamma_B$, from which we define the cavity-protection magnetic field value
\begin{equation}
    B_{\rm cp} = \frac{\hbar \Gamma_B}{\mu_B^*}.
\end{equation}
Using the Hikami-linewidth Eq. \eqref{eq:B_linewidth_approx_scaling}, we can derive an analytical expression for this quantity, 
\begin{equation}\label{eq:Bcp}
    B_{\rm cp} = \frac{\hbar}{e\xi_c^2}\left[ 1 + \frac{144\log (2)}{\pi}\, \left( \eta_{\rm dis} \frac{e \xi_c^2}{\hbar} \frac{\hbar \omega_{\rm qw} }{2\mu_B^*} \right)^2 \right]^{1/2} - \frac{\hbar}{e\xi_c^2}.
\end{equation}
Using some typical values such as $\xi_0 = 0.75\,$nm, $\xi_c = 5\,$nm and $L_z = 8\,$nm \cite{Unuma_doi:10.1063/1.1535733}, and considering the typical effective Bohr magneton $\mu_B^* \approx 0.86 \,$meV / T for the GaAs electron's effective mass, we obtain $\hbar \omega_{\rm qw} \approx 262\,$meV, and $B_{\rm cp} \approx 33.4\,$ T, which looks like a quite extreme value for the magnetic field.
Already for a slightly smaller value for the QW height fluctuations, for instance $\xi_0 = 0.3\,$ nm, we however obtain a much smaller and accessible result $B_{\rm cp} \approx 7.6\,$ T.
A complete plot of $B_{\rm cp}$ in unit of Tesla, as a function of the ISB frequency $\omega_{\rm qw}$ and the disorder correlation length $\xi_c$ in the typical range for THz-MIR ISB polaritons  is displayed in Fig. \ref{fig:fig10} for a fixed value of the QW length fluctuations $\xi_0 = 0.3\,$nm.

\begin{figure}
    \centering
    \includegraphics[width=\columnwidth]{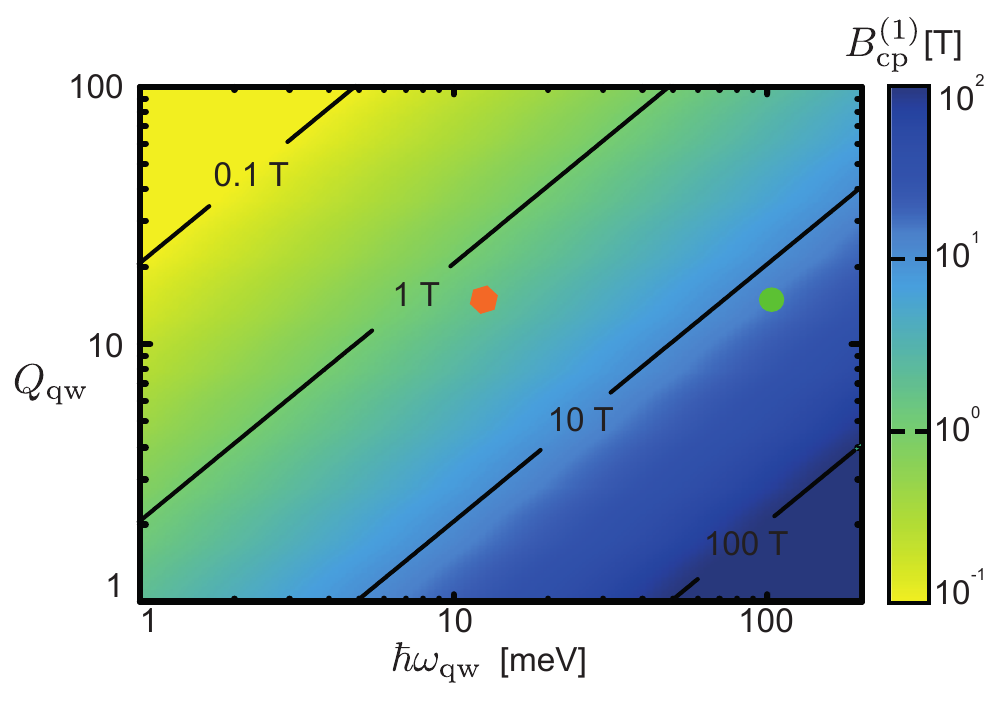}
    \caption{Log-scale contour-plot of the cavity-protection magnetic field $B_{\rm cp}^{(1)}$ as predicted by Eq.\eqref{eq:estimate_Bcp_exp} as a function of the intersubband frequency $\hbar\omega_{\rm qw}$ and the QW quality factor $Q_{\rm qw}$ (in log-scale). Here, $B_{\rm cp}^{(1)}$ is given in units of Tesla and $\hbar\omega_{\rm qw}$ in meV. The red hexagon and green circle mark the position of, respectively, typical THz~ \cite{Colombelli_THz_PhysRevLett.125.097403, Raab:19, Todorov_PhysRevLett.105.196402} and MIR~\cite{knorr2022intersubband} devices. }
    \label{fig:fig11}
\end{figure}

Since the values of $\xi_0$ and $\xi_c$ are hardly accessible to direct measurement, it is useful to derive an expression for $B_{\rm cp}$ as a function of  quantities that are directly accessible to experiments, like the central ISB frequency $\omega_{\rm qw}$ and the bare QW quality factor $Q_{\rm qw}$.
This can be achieved by expanding Eq. \eqref{eq:Bcp} to lowest order in the small disorder limit, and combining it with the Unuma-linewidth in Eq. \eqref{eq:disorder_strength_estimate_noB}
\begin{equation}\label{eq:estimate_Bcp_exp}
    B_{\rm cp}^{(1)} \approx \frac{16\log (2)}{\pi}\frac{1}{Q_{\rm qw}}\frac{\hbar \omega_{\rm qw} }{2\mu_B^*}.
\end{equation}
This expression is plotted in Fig. \ref{fig:fig11}, where we can clearly identify the regime of interest for THz-MIR devices, in the range of magnetic field between $B \sim 1-10\,$T. 
It is important to stress that these values of magnetic field are already within reach of current experiments, as also suggested by recent works on similar devices \cite{Scalari_PhysRevLett.93.237403, Uji_nature_2001, Feist_doi:10.1126/science.abl5818}.
Note that for the realistic parameters of disorder considered here, the actual values of $B_{\rm cp}^{(1)}$ and $B_0$ are quite close, with the approximated relation $B_{\rm cp}^{(1)} \approx 7/\sqrt{Q_{\rm qw}} B_0$.

As a final point, we need to address the second condition for cavity protection, regarding the Rabi frequency $\Omega_R$. Combining Eq. \eqref{eq:OmegaRcp} with Eq. \eqref{eq:estimate_Bcp_exp} we obtain
\begin{equation}
    \hbar \Omega_R = 2\mu_B^* B_{\rm cp}^{(1)} = \frac{16\log (2)}{\pi}\frac{1}{Q_{\rm qw}} \hbar \omega_{\rm qw}.
\end{equation}
For the range of quality factors estimated above, we obtain a range for the minimal Rabi frequency necessary to implement the magnetic-induced cavity protection in the order of 
\begin{equation}
    \frac{\Omega_R}{\omega_{\rm qw}} \sim 0.05-0.5.
\end{equation}
From this simple estimation it appears that the required Rabi frequency is not far from the ISB transition frequency, $\Omega_R \sim \omega_{\rm qw}$, pushing the system from the \emph{strong-coupling} towards the \emph{ultra-strong coupling} (USC) regime. Given  the complexities of the theoretical description of the USC regime \cite{nori_USC_review, Solano_USC_RevModPhys.91.025005} we have chosen to restrict here our attention to the strong coupling physics and we leave a specific investigation of the USC features to a future work.

As a final point, it is important to comment on the role of other processes that, in addition to the scattering on disorder potential, may contribute to the linewidth of the ISB~\cite{Unuma_doi:10.1063/1.1535733}.
While all decoherence channels stemming from {\em static} external potentials are tamed by the cavity protection mechanism discussed here for interface roughness, e.g. alloy disorder and ionized impurity, a special attention must be paid to phonon scattering processes. Longitudinal acoustic (LA) phonons typically have low frequencies, much lower than the intrinsic broadening of polaritons: as such, they can be considered as quasi-static and the virtually elastic LA phonon scattering thus fall in the same category of static potentials.

Longitudinal optical (LO) phonons have instead a much higher energy $E_{LO}$ on the order of a few ten meV in typical materials and their contribution to the linewidth is small but sizable in many cases, giving a lower bound to the achievable linewidth on the order of 1meV for transitions in the MIR according to~\cite{Unuma_doi:10.1063/1.1535733}. 
Still there exist regimes, e.g. THz ISB transitions with $\omega_{21}< E_{LO}$ and a low electron density such that $E_F<E_{LO}$ for which LO phonon emission is kinematically not allowed and the lower bound disappears.

\section{Conclusions}
\label{sec:conclusions}

In this work, we have proposed and characterized a strategy to dramatically improve the quality factor of intersubband polaritons in semiconductor-based devices. 

By applying a strong magnetic field perpendicular to the quantum well plane, the dominant Lorentzian broadening of the intersubband transition due to scattering of electrons onto interface roughness disorder turns into a Gaussian one due to the strong electron localization in the disorder potential. When strongly-coupled to the cavity mode, a cavity protection mechanism sets in which removes the Gaussian linewidth, leaving only the cavity contribution to the polariton linewidth. 

In combination with higher-$Q$ cavity configurations, our proposal has the potential to lead to polariton devices with unprecedented performances. Based on available experimental evidence, we anticipate that a narrower polariton linewidth will be a game-changing step in view of technological applications of intersubband polaritons, including nonlinear polariton devices~\cite{cominotti2021theory} and polariton lasing~\cite{Colombelli_Perspectives_PhysRevX.5.011031,knorr2022intersubband}. On a longer perspective, our proposal highlights intersubband polariton physics in the presence of magnetic fields as a novel arena where to study the interplay between light-matter interaction with quantum Hall physics~\cite{Atac_Science_doi:10.1126/science.1258595, Atac_PhysRevLett.120.057401, Atac_Nature_2019, Feist_doi:10.1126/science.abl5818,Ciuti_breakingTopoCavity_PhysRevB.104.155307, Rubio_doi:10.1126/science.abn5990, Rubio_PhysRevLett.123.047202}.


\acknowledgements
We are grateful to Giacomo Scalari, Ivan Amelio, Alberto Nardin and Jacopo Nespolo for fruitful discussions. We acknowledge financial support from the European Union FET-Open grant MIR-BOSE (737017).

\appendix

\section{Input-output theory of the cavity plasma Hamiltonian in the weak excitation regime}
\label{app:sec:in_out_th}

Consider the cavity-plasma Hamiltonian defined in Eq. \eqref{eq:Ham_cavity_plasma}.
Since we are looking only at its low energy excitations we can truncate the electronic Hilbert space keeping only the unperturbed Fermi sea state $| FS \rangle$ and its lowest single-electron-hole-pair excitations between the first two subbands, given by
\begin{equation}
| k_0 \, q_1 \rangle = c_{0\, k} c_{1 \, q}^{\dag} | FS \rangle,
\end{equation}
where $k \leq k_F$, with $k_F$ the Fermi momentum.
All the Fermionic operators in Eq. \eqref{eq:Ham_cavity_plasma} are then replaced by Pauli matrices, each of them representing a transition between the Fermi sea and one electron-hole state ($\hbar = 1$)
\begin{equation}\label{eq:Ham_cavity-plasma_spin}
\begin{split}
H_{\rm tot} & \approx \omega_c a^{\dag} a + \sum_{\vec{\lambda}} \left( \omega_{\rm qw} + \omega^{\parallel}_{\vec{\lambda}} \right) s_z^{\vec{\lambda}} 
\\
& + \frac{\Omega_R}{2}\left( a\cdot  \sum_{\vec{\lambda}} \Lambda_{\vec{\lambda}} s_+^{\vec{\lambda}}  + {\rm h.c.} \right).
\end{split}
\end{equation}
Here every couple $\vec{\lambda} = (k, k')$ identify a single two-level system and $s_z^{\vec{\lambda}} , s_-^{\vec{\lambda}}, s_+^{\vec{\lambda}} $ are the usual spin-$1/2$ operators. The range of the vectorial index $\vec{\lambda}$ is limited to the semi-rectangular area $(-\infty, \infty) \times [-k_F, k_F]$.

We then derive the usual quantum Langevin equations \cite{gardiner00} assuming a two-sided cavity and a generic bath for the ISB transitions. We have
\begin{equation}\label{eq:qLangevin_cavity}
    i\partial_t a = \left(\omega_c - i\frac{\gamma_c}{2} \right)a + \frac{\Omega_R}{2}\sum_{\vec{\lambda}} \Lambda_{\vec{\lambda}} s_-^{\vec{\lambda}} + \sqrt{\frac{\gamma}{2}}\alpha_{\rm in}(t) + b_{c},
\end{equation}
\begin{equation}\label{eq:qLangevin_s_-}
    \begin{split}
    i\partial_t s_-^{\vec{\lambda}} & = (\omega_{\rm qw} + \omega^{\parallel}_{\vec{\lambda}})s_-^{\vec{\lambda}} - \Omega_R \Lambda_{\vec{\lambda}} s_z^{\vec{\lambda}} a +\\
    &+ i 2\kappa s_z^{\vec{\lambda}} s_-^{\vec{\lambda}} - 2 s_z^{\vec{\lambda}} b_{\vec{\lambda}},
    \end{split}
\end{equation}
\begin{equation}\label{eq:qLangevin_s_z}
    \begin{split}
    i\partial_t s_z^{\vec{\lambda}} & = \frac{\Omega_R}{2}\left( \Lambda_{\vec{\lambda}}^*s_+^{\vec{\lambda}} a - \Lambda_{\vec{\lambda}} s_-^{\vec{\lambda}} a^{\dag} \right) + \\
    & - i2\kappa s_z^{\vec{\lambda}} - i\kappa  + b_{\vec{\lambda}} s_+^{\vec{\lambda}} - b_{\vec{\lambda}}^{\dag} s_-^{\vec{\lambda}}.
    \end{split}
\end{equation}
Here $\gamma_c$ is the cavity loss rate, $\kappa$ is the ISB transition relaxation rate, $b_c$ and $b_{\vec{\lambda}}$ are respectively the cavity and ISB transition quantum noise operators.
$\alpha_{\rm in}(t) = \alpha_{\rm in}^0 e^{-i\omega t}$ is the coherent input field , assumed to be monochromatic at a given frequency $\omega$.
The input $\alpha_{\rm in}$, reflected $\alpha_{\rm r}$, transmitted $\alpha_{\rm t}$ and cavity field $a$ are related by the following input-output formulas
\begin{equation}
    \begin{split}
    \alpha_{\rm r} & = \alpha_{\rm in} + \sqrt{\gamma_c/2} a,
    \\
    \alpha_{\rm t} & = \sqrt{\gamma_c/2}a.
    \end{split}
\end{equation}
We now consider the mean-field and weak driving regime. We then replace the operators $a$, $s_-^{\vec{\lambda}}$, $s_z^{\vec{\lambda}}$ in Eqs. \eqref{eq:qLangevin_cavity}-\eqref{eq:qLangevin_s_-}-\eqref{eq:qLangevin_s_z} with their respective expectation values (for simplicity in the notation we still keep the same notation for the operators and their mean value).
The quantum noise operators drop out since that at temperature $T=0$ they have zero mean value. Because of the weak drive we can also approximate $s_z^{\vec{\lambda}} \approx - 1/2$.

In the rotating frame at the input frequency $\omega$, the mean-field Langevin equations take the form of a driven dissipative system of coupled harmonic oscillators
\begin{equation}
    i\partial_t a = \left(\omega_c - \omega - i\frac{\gamma_c}{2} \right)a + \frac{\Omega_R}{2}\sum_{\vec{\lambda}} \Lambda_{\vec{\lambda}} s_-^{\vec{\lambda}} + \sqrt{\frac{\gamma}{2}}\alpha_{\rm in},
\end{equation}
\begin{equation}
    i\partial_t s_-^{\vec{\lambda}} = (\omega_{\rm qw} + \omega^{\parallel}_{\vec{\lambda}} -\omega )s_-^{\vec{\lambda}} + \frac{\Omega_R}{2} \Lambda_{\vec{\lambda}} a - i \kappa s_-^{\vec{\lambda}}.
\end{equation}
Solving these equations for the steady state, and defining the optical transmission through the cavity as
\begin{equation}
    T_c(\omega) = \frac{\alpha_{\rm t}}{\alpha_{\rm in}},
\end{equation}
we immediately arrive to Eq. \eqref{eq:cavity_transmission}.

\section{Electronic response and subband Green's function}
\label{app:sec:Green_fun}
The electronic optical response defined in Eq. \eqref{eq:electronic_Susceptibility_optical_response} can be rewritten within a Green's function formalism. We start from the initial formula
\begin{equation}
\begin{split}
&\chi_{\rm qw}(\omega )  =  - \sum_{k, k'\leq k_F} \frac{\left| \Lambda_{k, k'} \right|^2}{\omega - \omega_{\rm qw} - \omega_{k \, k'}^{\parallel} + i\kappa } 
\\
& =\int  \frac{d^2x\, d^2 y}{N_e}  \sum_{k, k'\leq k_F} \frac{ \braket{y | \varphi_{2 \, k}}\braket{\varphi_{2 \, k} | x} \braket{x | \varphi_{1 \, k'}}\braket{\varphi_{1 \, k'} | y} }{\omega - \omega_{\rm qw} - \omega_{k \, k'}^{\parallel} + i\kappa } 
\end{split}
\end{equation}
Using the Sokhatsky identity 
\begin{equation}
\lim_{\epsilon \rightarrow 0}{\rm Im} \left[\frac{1}{\omega - \omega_k + i\epsilon } \right] = -\pi \delta (\omega - \omega_k)\,
\end{equation}
we can write
\begin{equation}
\begin{split}
&\chi_{\rm qw}(\omega ) = 
\\
 & \int d\omega' d\omega'' \frac{d^2x d^2 y}{\pi^2 N_e}  \frac{{\rm Im}\left[ G_2(y,x, \omega')\right] {\rm Im}\left[ G_1(x,y, \omega'')\right] F(\omega '')}{\omega - (\omega' - \omega'') + i \kappa}
\end{split}
\end{equation}
where the Green's function is defined in the Lehman representation as
\begin{equation}
G_n(x,y, \omega) = \lim_{\epsilon \rightarrow 0}\sum_k \frac{ \braket{x | \varphi_{n \, k}}\braket{\varphi_{n \, k} | y}  }{\omega - \omega_{n, k} + i\epsilon},
\end{equation}
$\omega_{n, k} = \varepsilon^{\rm qw}_n/\hbar + \omega_n^{\parallel}(k)$ and $F(\omega )$ is the Fermi distribution. 
The QW optical spectral density is then given by
\begin{equation}
\begin{split}
&\rho_{\rm qw} (\omega )  = \lim_{\kappa \rightarrow 0}\frac{1}{\pi} {\rm Im}\left[ \chi_{\rm qw}(\omega ) \right] 
\\
&= \int d\omega' \frac{d^2x d^2 y}{N_e} \rho_2(y,x, \omega  - \omega ')  \rho_1 (x, y, \omega ') F(\omega ') 
\end{split}
\end{equation}
where
\begin{equation}\label{eq:local_DOS_single_band}
\rho_n (x, y, \omega ) = -\frac{1}{\pi} {\rm Im}\left[ G_n(x,y, \omega)\right]
\end{equation}
is the local spectral density of a single $n$-subband.

\section{Technical details on the numerical calculation}
\label{app:sec:numerical}
Here we consider the Hamiltonian as defined in Eq. \eqref{eq:Ham_in-plane_generic}, including the presence of the homogeneous magnetic field via the symmetric-gauge vector potential $\vec{A}(\vec{r}) = B/2(-y, x, 0)$,
\begin{equation}
H_n^{\parallel} = \frac{\left(\vec{p} - e \vec{A}(\vec{r})\right)^{\, 2}}{2m^* } + \delta U_n(\vec{r})\,. 
\end{equation}
Here, the disorder potential is
\begin{equation}
\delta U_n(\vec{r}) = \partial_L \varepsilon_n(L_{\rm qw}) \cdot \delta L(\vec{r}),
\end{equation}
in terms of the energy  $\varepsilon_n(L_{\rm qw})$ of the electronic state trapped in the QW in the $z$-direction.
The QW interface roughness $\delta L$ is a random variable, in general with non-trivial spatial correlations. We can thus write
\begin{equation}
\overline{\delta L(\vec{r}) \delta L(0)} = \xi_0^2 C(\vec{r}),
\end{equation}
where $\xi_0$ quantifies the magnitude of the fluctuations of the QW thickness and $C(\vec{r})$ is an arbitrary correlation function normalised to have $C(0)=1/(2\pi )$.
We can then introduce the adimensional random variable $\Delta(\vec{r}) = \delta L(\vec{r})/\xi_0$.

We then rescale the position and the momenta as
\begin{align}
\vec{r} \longmapsto \vec{r} \cdot l_{\rm qw} && \vec{p} \longmapsto \vec{p} \cdot \hbar/l_{\rm qw}
\end{align}
in terms of an equivalent harmonic-oscillator length within the QW
\begin{equation}
l_{\rm qw} = \left( \frac{\hbar}{ m^* \omega_{\rm qw}} \right)^{1/2}.
\end{equation}
Defining $\vec{a} = (-y, x)/2$, and considering the usual cyclotron frequency as $\omega_B = eB/m^*$, we have
\begin{equation}\label{eq:rescaled_H_n}
H_n^{\parallel} = \hbar \omega_{\rm qw}\left[ \frac{\left( \vec{p} - \eta_B\vec{a} \right)^2}{2}  + g_n \eta_{\rm dis} \Delta (\vec{r}) \right],
\end{equation}
where we introduced the two adimensional energy magnitudes
\begin{align}
\eta_B = \frac{\omega_B}{\omega_{\rm qw}} && \eta_{\rm dis} = \frac{\partial_L \varepsilon_1(L_{\rm qw}) \xi_0}{\hbar \omega_{\rm qw}},
\end{align}
and the numerical coefficients
\begin{equation}
    g_n = \frac{\partial_L \varepsilon_n(L_{\rm qw})}{\partial_L \varepsilon_1(L_{\rm qw})}.
\end{equation}
In the specific case of an infinite box potential $g_n = n^2$, while in the harmonic oscillator case $g_n = n - 1/2$, with $n=1,2 \ldots$ .

In our numerical calculations, the adimensional disorder is generated by sampling $\Delta (\vec{r})$ from a Gaussian distribution with unit variance at every position $\vec{r}$. We then take its Fourier transform and we impose a Gaussian cut-off
\begin{equation}
\tilde{\Delta} (\vec{k} ) \longmapsto \frac{\xi_c}{\sqrt{2}} e^{-\xi_c^2 k^2/8} \tilde{\Delta}(\vec{k}).
\end{equation}
After transforming back to real-space, we are left with the desired correlator
\begin{equation}\label{app:eq:gaussian_correlator}
    \overline{\Delta (\vec{r}) \Delta (\vec{r}^{\, '})} = \frac{e^{-|\vec{r}- \vec{r}^{\, '}|^2/\xi_c^2}}{2\pi}.
\end{equation}

We then proceed to diagonalize the rescaled Hamiltonian Eq. \eqref{eq:rescaled_H_n} on a given basis. 
When the magnetic field is not present $B=0$ we consider a finite box of lengths $L_x, L_y$ whose basis wave functions are
\begin{equation}
    \varphi_{n_x, n_y}(\vec{r}) = \frac{2}{\sqrt{L_xL_y}}\sin ( \frac{\pi n_x}{L_x}x ) \sin ( \frac{\pi n_y}{L_y}y ),
\end{equation}
with $n_x,n_y=1,2\ldots$. Same results are obtained considering a periodic system and considering a planewave eigenbasis.

In the presence of a finite magnetic field $B\neq 0$, we first need to fix a gauge for the vector potential.
For instance, in the symmetric gauge where $\vec{a} = (-y, x)/2$.
In order to simulate an infinite system with open boundary conditions we use the basis set
\begin{equation}
    \varphi_{\ell k}(\vec{r}) = \frac{1}{\sqrt{2\pi}l_B} \sqrt{\frac{\ell !}{k !}}\xi^{k-\ell} e^{-|\xi|^2/2}L_{\ell}^{k-\ell}(|\xi|^2),
\end{equation}
where $\xi = (x+iy)/(\sqrt{2}l_B)$ and $L_{\ell}^{\alpha}(x)$ are the generalised Laguerre polynomials.
Since these states are concentric circles, centered in the origin, we include as many states as possible to fill the spatial extension of the system, cutting the basis right before touching the border of the numerical space grid. If the total extension of all basis states $L_{\rm basis}$ is such that $L_{\rm basis} \gg l_B, \xi_c$, averaging over many realisations is then approximately similar to consider a very large system, where every realisation is a smaller patch of the whole system. 

In order to compute the matrix elements on a given basis we consider a two dimensional spatial grid $N_x \times N_y$, where we typically use a number of grid points in the range $N_x = N_y = 100-150$, with a grid step in the range $\Delta r/L_z = 0.1-0.5$, dependently from the chosen values of $\xi_c$ and $l_B$.
All final results are typically averaged over $N_{\rm dis}\sim 20-100$ realisations.

Notice that we do not include the edge states in our calculation. This is motivated by the quantitative smallness of their contribution, which is completely negligible in an extended system. Indeed in a very large system the edge modes represents a much smaller fraction of the whole system, and, since the dipole moment along $z$ is the same for every electron, localised or non-localised, their contribution is negligible with respect to the total number of bulk states.

\section{Lowest Landau level disordered density of state}
\label{app:sec:disordered_densityLLL}
In this section we briefly review the calculation of the disordered density of states in the lowest Landau level (LLL) contained in \cite{hikami_1985_:jpa-00210150}. For simplicity we work in adimensional units where $\omega_{\rm qw} = 1$.

The calculation starts from the identity in Eq. \eqref{eq:local_DOS_single_band}.
What we need to calculate is the Green's function for the disordered Schr\"odinger equation, projected in the LLL and averaged over the disorder.
To do so we introduce a quantum field theory representation of the Schr\"odinger Green's function in terms of complex scalar field path integral \cite{Zee:706825}
\begin{equation}
    \begin{split}
        &G(x,y, \omega) = \braket{x|\frac{1}{\omega - H} |y} =
        \\
        & \frac{1}{i}\frac{\int \mathcal{D}\phi \mathcal{D}\phi^{*} e^{i\int d^2 x' \left[ \phi(x')^* (\omega - H)\phi(x') \right]}\phi(x)\phi^*(y)}{\int \mathcal{D}\phi \mathcal{D}\phi^{*} e^{i\int d^2 x' \left[ \phi(x')^* (\omega - H)\phi(x') \right]}}.
    \end{split}
\end{equation}
We then use the relation between path integral and functional determinant to transform the Bosonic path integral in the denominator into a Grassmannian path integral 
following the so-called super-symmetric approach \cite{BREZIN198424}
\begin{equation}
    \begin{split}
        & \frac{1}{\int \mathcal{D}\phi \mathcal{D}\phi^{*} e^{i\int d^2 x' \left[ \phi(x')^* (\omega - H)\phi(x') \right]}} = \det \left[ \omega - H \right] =
        \\
        & = \int \mathcal{D}\eta \mathcal{D}\bar{\eta} e^{i\int d^2 x' \left[ \eta(x') (\omega - H)\bar{\eta}(x') \right]}.
    \end{split}
\end{equation}
Here $\eta(x)$ is an anticommuting Grassmann field.
Considering our Hamiltonian composed by kinetic energy $\hat{T}$ and disorder potential $\delta U$ terms 
\begin{equation}
    \begin{split}
        & \braket{x|\frac{1}{\omega - H} |y} = \frac{1}{i}\int \mathcal{D}\phi \mathcal{D}\phi^{*} \mathcal{D}\eta \mathcal{D}\bar{\eta} \, \phi(x)\phi^*(y) \times
        \\
        & \times \exp \left[ i \int d^2x'\left[ \phi^* \left( \omega - \hat{T} \right)\phi + \eta \left( \omega - \hat{T} \right) \bar{\eta} \right] \right] \times
        \\
        & \times \exp \left[ -i \int d^2x' \delta U(x') \left( \phi^* \phi + \eta \bar{\eta} \right) \right].
    \end{split}
\end{equation}

We now take the average over the disorder. Since we are considering a Gaussian disorder, we can safely apply the second cumulant expansion, 
\begin{equation}
\begin{split}
    &\overline{\exp\left[-i\int d^2x' \delta U(x') f(x') \right]} = 
    \\
    & = \exp \left[ - 1/2\int d^2x_1 d^2x_2 \overline{\delta U(x_1) \delta U (x_2)}f(x_1) f(x_2) \right].
\end{split}
\end{equation}
where $\overline{\cdots}$ indicates the average over many disorder realisations.
We then introduce the disorder correlator as $\overline{\delta U(x_1) \delta U (x_2)} = C(x_1 - x_2)$.
In the case of infinite range correlations ($\xi_c \rightarrow \infty$) we have $C(x_1-x_2) = C_0$, and
the averaged Green's function reads
\begin{equation}\label{eq:pathInt_rep_avgGfun_infiniteCorr}
    \begin{split}
        & \overline{\braket{x|\frac{1}{\omega - H} |y} } = \frac{1}{i}\int \mathcal{D}\phi \mathcal{D}\phi^{*} \mathcal{D}\eta \mathcal{D}\bar{\eta} \, \phi(x)\phi^*(y) \times
        \\
        & \times \exp \left[ i \int d^2x'\left[ \phi^* \left( \omega - \hat{T} \right)\phi + \eta \left( \omega - \hat{T} \right) \bar{\eta} \right] \right] \times
        \\
        & \times \exp \left[ - \frac{C_0}{2} \left(\int d^2x' \left( \phi^*(x') \phi(x') + \eta(x') \bar{\eta}(x') \right) \right)^2 \right].
    \end{split}
\end{equation}
We now project the fields over the LLL considering $\phi(x) = \sum_k \varphi_{0\, k}(x) a_k$ and $\eta(x) = \sum_k \varphi_{0, \, k} b_k$, where $\varphi_{\ell, k}$ is the wave function of the $k$-th state in the $\ell$-th Landau level.
We also insert in the integral an identity operator using a delta function and we shift $\omega \mapsto \omega -\omega_B/2$.

The Green's function then reads
\begin{equation}\label{eq:avg_gFun_integral}
\overline{\braket{x|\frac{1}{\omega - H} |y} } = \frac{1}{i} \int_{-\infty}^{+\infty} d \sigma \, e^{- \frac{C_0}{2}\sigma^2 + i\omega\sigma} D(x,y,\sigma)
\end{equation}
with
\begin{equation}
    D(x,y,\sigma ) = \sum_k \varphi_{0\, k}(x) \varphi^*_{0\, k}(y) I_k(\sigma ),
\end{equation}
and
\begin{equation}
    \begin{split}
        & I_k(\sigma ) = \int \Pi_{k'} \, d^2a_{k'} d^2 b_{k'} \, a_k^* a_k \, \delta \left(\sigma - \sum_k (a_k^* a_k + b^*_k b_k) \right)
        \\
        & = \lim_{\varepsilon \rightarrow 0} \int \frac{dt}{2\pi}  \int \Pi_{k'} \, d^2a_{k'} d^2 b_{k'} \, a_k^* a_k e^{i t \sigma - i(t-i\varepsilon)\sum_k (a_k^* a_k + b^*_k b_k) }
        \\
        & = \int \frac{dt}{2\pi i}\frac{e^{it\sigma}}{t - i\varepsilon} = \Theta (\sigma )
    \end{split}
\end{equation}
in terms of the step function $\Theta(\sigma)$.

To get the density of states, we need to set $x=y$, integrate over the whole area $A$ of the system and finally normalise over the total electron number. Since $\sum_k |\varphi_{0\, k}(x)|^2 = 1/(2\pi l_B^2)$ we get a factor $A/(2\pi l_B^2)$ from the spatial integration. In a finite system the total number of states in the LLL is $N_{LLL}\sim A/(2\pi l_B^2)$ \cite{Huckestein_1995_RevModPhys.67.357}. 
Completing the square, performing the Gaussian integral in Eq. \eqref{eq:avg_gFun_integral} and using Eq. \eqref{eq:local_DOS_single_band} we finally find the density of states for the LLL in the infinite correlation-length approximation
\begin{equation}
\rho_{LLL}^{\infty}(\omega ) \approx \frac{1}{\eta_{\rm dis} }e^{-\frac{\pi \omega^2}{\eta_{\rm dis}^2}},
\end{equation}
where we considered that $C_0 = \eta_{\rm dis}^2/(2\pi)$.
From this distribution we extract the Hikami linewidth in the infinite correlation-length case $\Gamma_{B\, {\rm H}}^{\infty}$ defined in Eq.\eqref{eq:hikami_LW_infinite}.

In order to include information about the finite correlation length, we expand the correlator in series as $C(x) \approx C_0(1 - x^2/\xi_c^2)$.
In Eq. \eqref{eq:pathInt_rep_avgGfun_infiniteCorr} we need to add to the argument of the exponential in the last line the following term
\begin{equation}
\begin{split}
    & \frac{C_0}{2}\int d^2 x_1 d^2 x_2 \frac{|x_1 - x_2|^2}{\xi_c^2}\left( \phi^*(x_1) \phi(x_1) + \eta(x_1) \bar{\eta}(x_1) \right) \times 
    \\
    & \times \left( \phi^*(x_2) \phi(x_2) + \eta(x_2) \bar{\eta}(x_2) \right)
    \\
    & \approx \frac{2l_B^2}{\xi_c^2}\left[ \sum_k (a_k^* a_k + b^*_k b_k). \right]^2.
\end{split}
\end{equation}
We see that this term contributes only by a shift of $C_0$ which is replaced by $C_0-2C_0l_B^2/\xi_c^2$.
We then re-sum all the terms similar to this one from the higher order correlator expansion considering that $C_0-2C_0l_B^2/\xi_c^2 \approx C_0/(1+2l_B^2/\xi_c^2)$.

In this way, we arrive to the LLL density of states in the case of a finite correlated disorder potential
\begin{equation}
    \rho_{LLL}(\omega) \approx \frac{1}{\sqrt{2\pi}\gamma_{LLL}}e^{-\frac{\omega^2}{2\gamma_{LLL}^2}},
\end{equation}
where
\begin{equation}
    \gamma_{LLL} = \frac{\eta_{\rm dis}\xi_c/l_B}{\sqrt{2\pi \left( \xi_c^2/l_B^2 + 2 \right)}}.
\end{equation}
The FWHM reported in Eq. \eqref{eq:LLL_FWHM} is then equal to $\Gamma_{LLL} = 2\sqrt{2\log (2)}\, \gamma_{LLL}$.

\section{Rabi splitting and polaritonic linewidth}
\label{app:pol_linewidth}

Here we summarise the results contained in \cite{Diniz_CavityProtection_PhysRevA.84.063810} regarding the cavity protection effect.

The poles of the cavity transmission Eq. \eqref{eq:cavity_transmission} give the system's resonant frequencies and their linewidth, through the equation $T_c^{-1}(\omega) = 0$.
 
In a perfectly clean system or when the disorder potential is identical for the two subbands, 
we have $\Lambda_{k\, k'}= \delta_{k k'}$. We thus obtain the standard cavity transmission modified by the coupling to the QW
\begin{equation}
T_c^{\rm clean}(\omega) = - \frac{\gamma_c/2}{\omega - \omega_c + i \frac{\gamma_c}{2} + \frac{\Omega_R^2}{4}\chi_{\rm clean}(\omega) },
\end{equation}
where
\begin{equation}
\chi_{\rm clean}(\omega) = -\frac{1}{\omega - \omega_{\rm qw} + i\frac{\kappa}{2}}
\end{equation}
is the resonant response of the QW.
Assuming the light-matter resonance condition $\omega_c \sim \omega_{\rm qw}$ in the  strong coupling regime $\Omega_R^2/(\gamma_c \kappa)\gg 1$, we have that the cavity transmission has two resonant peaks, corresponding with the polaritonic frequencies $\omega_{\pm } \approx \omega_c \pm \Omega_R/2$. The linewidth of these two peaks is the same and is given by the average between the cavity and the QW linewidths $\Gamma \approx (\gamma_c + \kappa)/2$.

In a situation of small disorder we still expect to have the two polariton peaks around the frequencies $\omega_{\pm}$ but with a modified linewidth.
In order to derive this new linewidth we split the QW optical response in real and imaginary part $\chi(\omega) = \chi^R(\omega) + i \chi^I(\omega)$, considering that $\chi^I = \pi \rho (\omega)$. We also work in a rotating frame, such that $\omega-\omega_c \longmapsto \omega$, with $\omega_c = \omega_{\rm qw}$.
The poles of the cavity transmission are then given by
\begin{equation}\label{eq:cavity_plasmon_poles_equation}
\omega +\frac{\Omega_R^2}{4}\chi^R(\omega) + \frac{i}{2}\left( \gamma_c + \frac{\pi}{2} \Omega_R^2 \rho(\omega ) \right) = 0.
\end{equation}
Assuming the strong coupling regime, where $\Omega_R$ is much larger than the linewidth of $\rho(\omega)$
\begin{equation}\label{eq:expansion_Chi_real}
    \chi^R(\omega) = - \int d\omega ' \frac{\rho (\omega ' )}{\omega - \omega ' } \approx -\frac{1}{\omega},
\end{equation}
and we approximate 
\begin{equation}
    \rho(\omega ) \approx \rho(\omega_{\pm} ) \,,
\end{equation}
From these relations we immediately obtain the approximated pole equation, valid only in the proximity of $\omega \sim \omega_{\pm}$
\begin{equation}
\omega^2 - \frac{\Omega_R^2}{4} +
 + i\left( \frac{\gamma_c}{2} + \frac{\pi}{4}\Omega_R^2 \rho(\omega_{\pm} ) \right)\omega  = 0 .
\end{equation}
Solving it we find
\begin{equation}
    \omega = \pm \frac{\Omega_R}{2}\sqrt{1 - \frac{\Gamma_{\pm}^2}{2\Omega_R^2}} - i \frac{\Gamma_{\pm}}{2},
\end{equation}
where the polaritonic linewidth is given by
\begin{equation}
    \Gamma_{\pm} = \frac{\gamma_c + \frac{\pi}{2}\Omega_{R}^2\rho(\omega_{\pm})}{2}
\end{equation}

\bibliographystyle{mybibstyle}
 
\bibliography{references}

\end{document}